\documentstyle[12pt,aasms4]{article}

\received{}
\accepted{}
\journalid{}{}
\articleid{}{}

\lefthead{Brunner et al.}
\righthead{Photometric-Redshifts}

\newcommand{\ie}{{\em i.e.\ }}
\newcommand{\eg}{{\em e.g.,\ }}
\newcommand{\etal}{{\em et al.\ }}
\newcommand{\cf}{{\em cf.\ }}

\newcommand{\UBRI}{{$U, B, R, \&\ I $}}
\newcommand{\UBR}{{$U, B, \&\ R $}}

\newcommand{\ubr}{{\em U B R\ }}

\newcommand{\U}{{\em U\ }}
\newcommand{\B}{{\em B\ }}
\newcommand{\V}{{\em V\ }}
\newcommand{\R}{{\em R\ }}
\newcommand{\I}{{\em I\ }}

\newcommand{\SqDegree}{Sq. Degree}
\newcommand{\SqArcminute}{Sq. Arcminute}

\newcommand{\HST}{{\it HST\ }}

\begin{document}

\title{The Statistical Approach to Quantifying Galaxy Evolution}

\author{Robert J. Brunner\altaffilmark{1,2}, 
Andrew J. Connolly\altaffilmark{1}, 
and Alexander S. Szalay}
\affil{\{rbrunner,ajc,szalay\}@pha.jhu.edu\\
Department of Physics \& Astronomy, The Johns Hopkins University, 
Baltimore, MD 21218}

\altaffiltext{1}{Visiting Astronomer, Kitt Peak National Observatory, 
National Optical Astronomy Observatories, which is operated by the
Association of Universities for Research in Astronomy, Inc. (AURA)
under cooperative agreement with the National Science Foundation.}
\altaffiltext{2}{Current Address: rb@astro.caltech.edu,
Department of Astronomy, California Institute of Technology, MC 105-24, 
Pasadena, CA 91125}

\begin{abstract}

Studies of the distribution and evolution of galaxies are of
fundamental importance to modern cosmology; these studies, however,
are hampered by the complexity of the competing effects of spectral
and density evolution. Constructing a spectroscopic sample that is
able to unambiguously disentangle these processes is currently
excessively prohibitive due to the observational requirements. This
paper extends and applies an alternative approach that relies on
statistical estimates for both distance ($z$) and spectral type to
a deep multi-band dataset that was obtained for this exact purpose.

These statistical estimates are extracted directly from the
photometric data by capitalizing on the inherent relationships between
flux, redshift, and spectral type.  These relationships are
encapsulated in the empirical photometric redshift relation which we
extend to $z \approx 1.2$, with an intrinsic dispersion of $\delta z
\sim 0.06$. We also develop realistic estimates for the photometric
redshift error for individual objects, and introduce the utilization
of the galaxy ensemble as a tool for quantifying both a cosmological
parameter and its measured error. We present deep, multi-band, optical
number counts as a demonstration of the integrity of our sample. Using
the photometric redshift and the corresponding redshift error, we can
divide our data into different redshift intervals and spectral
types. As an example application, we present the number redshift
distribution as a function of spectral type.

\end{abstract}


\section{Introduction}
With the advent of the Hubble Deep Field (\cite{williams96}), which
imaged objects that are approximately 100 times fainter than the limit
of ground based spectrographs, deriving redshifts from broadband
photometry has undergone a renaissance. However, the utility of
deriving galaxy redshifts from photometric data has long been known
(\cite{baum62,koo85,loh86}). The majority of the more recently
promoted redshift estimation techniques rely on fitting template
spectral energy distributions to the observed galaxy magnitudes
(\cite{lanzetta96,gwyn96,sawicki96,mobasher96}).  An alternative,
empirical, approach was developed by Connolly \etal (1995), in which
the redshift is estimated directly from the broadband
magnitudes. Utilizing photographic data, they were able to estimate a
redshift out to $z \sim 0.5$ with a measured dispersion of $\delta z <
0.05$. The uncertainties in that result were dominated by the
photometric errors, and simulations indicated that with improved
photometry, the dispersion within the relationship could be
significantly reduced.

Preliminary results (\cite{brunner97}) from these data indicate that
this empirical approach can recover redshifts from broadband
magnitudes with a measured intrinsic dispersion of $\delta z \approx
0.02$ to $z \sim 0.4$, which is remarkably close to the asymptotic
intrinsic dispersions ($\delta z \approx 0.016$ for $z < 0.4$) that
were previously predicted from simulated distributions of galaxy
colors. The rest of this paper extends this published analysis to
higher redshifts and provides a framework for quantifying galaxy
evolution. Specifically, we discuss the observations and data
reduction in Section~\ref{dataSection}. In Section~\ref{numCounts} we
present the deep, multi-band number counts from this data. The
photometric redshift and corresponding error estimates are presented
in section~\ref{photoZ}; and the classification of the data by
spectral type data is detailed in Section~\ref{sedType}.  We present
the number redshift distribution in Section~\ref{nOfZApp} as a simple
application of the statistical approach. We conclude this paper with a
discussion of the general technique and its applicability.

\section{Data\label{dataSection}}

\subsection{Photometric Observations}

The photometric data presented in this paper are located in the
intersection between the \HST 5096 field and the CFRS 14 hour field
(\ie the Groth Strip), covering approximately 0.054 \SqDegree. All of
the photometric data were obtained using the Prime Focus CCD (PFCCD)
camera on the Mayall 4 meter telescope at Kitt Peak National
Observatory (KPNO). The observations were performed on the nights of
March 31 - April 3, 1995, March 18 - 20, 1996, and May 14 - 16,
1996. The PFCCD uses the T2KB CCD, a $2048^2$ Tektronix CCD with 24
micron pixel scale, which at $f/2.8$ in the 4 meter results in a scale
of $0.47 \arcsec/$pixel and a field of view of $\approx 16.0\arcmin
\times 16.0\arcmin$. The T2KB has a measured RMS read noise of $4
e^{-}/$pixel (\cite{massey95}), and has a high charge transfer
efficiency as well as reasonable throughput in the
ultra-violet.

All observations were made through the broadband filters: \UBRI,
commonly referred to at Kitt Peak as the ``Harris Set'' (\cf
Figure~\ref{effectiveFilters}). These filters were chosen due to their
large spectral coverage, and are commonly used in deep, sky limited
broadband imaging. The total integration times for each filter are
listed along with the corresponding magnitude limits in
Table~\ref{limits}.

\placetable{limits}
\placefigure{effectiveFilters}

During the course of the program observations, the standard set of
calibration data: bias frames, dark frames, and flat fields, were
taken. These data are necessary to both properly remove the
instrumental signature from the data, as well as to provide real-time
integrity tests for both the CCD and the electronics.  Multiple flat
field images were taken at both the beginning and end of each night by
illuminating a white screen mounted inside the dome. A visual
inspection of the illumination pattern uncovered an irregularity which
was manifested in the flat field images themselves. This required the
construction of illumination correction images during the reduction
process. For all of our program data, we utilized a short scan of
fifty rows.

One final concern in using the PFCCD on the Mayall 4 meter was
maintaining a consistent focus over the image. The focus solution
varied throughout the night due to temperature fluctuations, and our
adopted solution was to monitor the PSF within our images. The focus
value for the current filter was chosen to optimize the PSF at
approximately one third of the radial distance outward from the center
of the image. Focus offsets that were determined at the beginning of
each night were then applied when switching between the different
program filters.

\subsection{Photometric Data Reduction}

The photometric data were reduced in the standard fashion and are
extensively detailed elsewhere (\cite{myThesis}); however, due to the
relevance of the accuracy of the photometric measurements, we present
a brief overview of the process. The bias pattern, which was very
stable, was removed by using a global bias frame, which was
constructed by combining all of the acceptable bias frames for a given
run. Variations in the bias level were removed using the overscan
region for each individual program image. During each run, the dark
current was examined and found to be both stable and uniform; and, as
a result, we applied no corrections for dark current in any of our
data frames. The small scale pixel to pixel variations were removed
using dome flats, and the residual large scale gradients were removed
using illumination corrections which were created from smoothed
super-sky flat images. A fringe correction image was constructed from
the \I band sky flat to remove any fringing due to OH line emission.

Cosmic rays (radiation events) were removed both by looking for
deviant pixels, and by utilizing a pixel rejection strategy during the
image stacking phase. Cosmetic defects were corrected by linearly
interpolating over them. Any charge depletions introduced into the
images by the CCD electronics were corrected (\cf Brunner 1997),
reducing any remaining variations to a few tenths of a percent.

Before combining the images, the geometric distortions were mapped
using a third order polynomial with cross terms fit to nine reference
stars. The images were transformed to the reference image, and stacked
using signal-to-noise weighting for each separate run. The final
stacked images for each of the three runs were also registered,
stacked, and trimmed in a similar manner. The final image section for
the deep stacked region is 1641 rows by 1943 columns, or $\approx
0.054$ \SqDegree.

\subsection{Photometric Calibration}

All photometric calibrations were done using the April 1995 dataset
which included published standard star fields (\cite{landolt92}). Once
the calibration frames were reduced, a curve of growth was generated
for each standard star in concentric apertures of diameters from
$4\arcsec$ to $20\arcsec$. All standard stars for which the curve of
growth converged were used to determine the photometric solution. A
linear regression on the published stellar magnitude, the instrumental
magnitude, the airmass, and a color term was performed, and the result
translated to a one second standard exposure. We transformed our
magnitude system to the AB system (\cite{okeGunn83}) using published
transformations (\cite{fukugita95}).

During the registration and stacking phase of the data reduction, both
photometric and non-photometric frames were combined. As a result, the
photometric calibration required a direct cross calibration between
the stacked image and a photometric calibrated reference image. This
cross calibration was done using a large number of high signal to
noise reference stars in order to derive the photometric relationship
between the two frames. The dispersion in the cross calibration
relationship between the stacked and reference frames was a few
hundredths in each band, and our overall intrinsic photometric
accuracy was better than two percent.

\subsection{Source Detection and Photometry}

Source detection and photometry were performed using SExtractor
version 2.0.8 (\cite{bertin96}) with the appropriate correction for
the background estimation bug applied (\cite{bertin98}). SExtractor
was chosen for its ability to detect objects in one image and analyze
the corresponding pixels in a separate image; which, when applied
uniformly to multi-band data, generates a matched aperture
dataset. Our detection image was constructed from the \UBRI\ images
using a $\chi^2$ process (\cite{szalay98}). Briefly, this process
involves convolving each input image with a Gaussian kernel matched to
the seeing. The convolved images were squared, and normalized so that
they had zero mean and unit variance. The four processed images
(corresponding to the original \UBRI\ images) were coadded, forming
the $\chi^2$ detection image. A histogram of the pixel distribution in
the $\chi^2$ image was created, and compared to a $\chi^2$ function
with four degrees of freedom (which corresponds to the sky pixel
distribution). The ``object'' pixel distribution was generated by
subtracting the ``sky'' pixel distribution from the actual pixel
distribution. The Bayesian detection threshold was set equivalent to
the intersection of the ``sky'' and ``object'' distributions (\ie where
the object pixel flux becomes dominant). To convert this empirical
threshold for use with SExtractor, we had to scale the threshold
(which is a flux per pixel value) into a surface brightness threshold
(which is in magnitudes per square arcsecond).

\subsection{Completeness and Photometric Limits\label{catalogLimits}}

Detecting and analyzing faint objects in an image is hindered by many
different effects, including object superposition, noise fluctuations,
as well as redshift and evolutionary effects. These factors can
prevent objects from being detected (confidence), introduce false
detections (contamination), and affect the accuracy of the photometric
measurements. The standard method for quantifying the reduction in the
detection efficiency is through a heavy utilization of simulations:
either through the simulation of an entire image, or through the
addition of simulated or real images to the actual object frame. In
order to determine the completeness limits in a model independent
manner, we adopted the latter approach.

We, therefore, added one hundred artificially generated galaxies to
the final stacked image. The magnitudes of the artificial galaxies
were drawn uniformly from a one half magnitude interval (\ie $25.0
\leq U < 25.5$). The new image was processed in an identical manner
to the original image, and the resultant catalog was cross-identified
with the input artificial galaxy catalog using both a proximity and
magnitude match criteria. This process was repeated ten times for each
half-magnitude bin in each filter over the relevant range of
magnitudes, and the mean and one sigma deviation were extracted using
a high/low rejection for each half-magnitude bin. These data were used
to determine the completeness correction curves (\cf
Figure~\ref{completeness}) by interpolating between the data points
with a second order polynomial. From the interpolating function, both
a $90\%$ and $50\%$ completeness limits in all four bands were
measured.

\placefigure{completeness}

Rather than relying on completeness limits, some analysis techniques
require precise photometric measurements. In order to satisfy these
conditions, we determined both the $2\%$ and $10\%$ photometric error
magnitude limits. These limits for all four bands were calculated by
scanning though the master catalog for all valid detections which had
a measured photometric error that was approximately the same as the
target photometric error ($0.1$ magnitudes for $10\%$ photometry and
$0.02$ magnitudes for $2\%$ photometry). The means of the magnitudes
for all such selected galaxies were determined, and these values were
designated as the magnitudes at which the photometric error reached
the target error (\cf Table~\ref{limits}).

\subsection{Astrometry}

Traditionally, astrometry is performed by identifying calibrated stars
within the image of interest.  Often these stars are selected from the
\HST Guide Star Catalog (GSC) (\cite{lasker88}). Within our stacked
image, however, we could not reliably utilize the GSC calibrated
stars, since there were not enough unsaturated GSC stars in our final
stacked image. Fortunately, we were able to obtain an early release of
the \HST Guide Star Catalog II (\cite{lasker96}), which, although it
is currently less precise than the original guide star catalog, had
sufficient calibration candidates to properly determine an astrometric
solution. The residuals of the final geometric transformation to the
GSCII for the reference stars were all less than $0.15$ pixels, or
equivalently, less than $0.07\arcsec$.

\subsection{Spectroscopic Cross-identification\label{spectroXID}}

To derive an empirical photometric redshift relation, a sample of
calibrating redshifts are required. Two spectroscopic surveys: the
Canada--France Redshift Survey (CFRS,~\cite{lilly95}), and the Deep
Extragalactic Evolutionary Probe (DEEP,~\cite{mould93}), have both
obtained spectroscopic redshifts for objects within our catalog.

The CFRS spectroscopic targets were selected from a complementary CCD
imaging survey to $I_{AB} \leq 22.5$. The spectroscopic observations
were done using the Canada-France-Hawaii 3.5 meter telescope. The DEEP
project is a multi-institution collaboration, in which we are
participating, that is currently using the low resolution imaging
spectrograph (LRIS) on the Keck II 10 meter telescope to obtain
spectra for objects to $I \sim 24.0$. DEEP spectroscopic targets are
predominantly selected from images obtained with the repaired \HST
wide field and planetary camera (WFPC2). The remaining spectroscopic
targets are selected from existing ground based imaging.

The object match-up procedure with each spectroscopic catalog was done
using a growing annulus technique where the angular distance $\psi$
was determined using the formula $\psi = \arccos(\sin(\phi_{s})
\sin(\phi_{p}) + \cos(\phi_{s}) \cos(\phi_{p}) \cos(\theta_{s} -
\theta_{p}))$, where the angles ($\theta \equiv$ Declination, $\phi
\equiv$ Right Ascension) have been properly converted into radians,
and the subscript s refers to spectroscopic target and the subscript p
refers to photometric object. A magnitude restriction was also
utilized to prevent improper identifications due to poor relative
astrometry, blended objects, or false detections. As both
spectroscopic surveys had \I band magnitudes for the spectroscopic
targets, the magnitude test was done using \I magnitudes.

\subsection{Star Galaxy Separation}

In an effort to simplify the star--galaxy separation effort, we
measured several different aperture magnitudes in addition to the
``Kron'' magnitude we use for the subsequent analysis. Of primary
interest is the $2.5$ pixel diameter aperture magnitude (matched to the
median point spread function), which provides a measurement of the
core flux of an object. The ratio of the core flux to the total object
flux should be larger for the bright stellar objects than for
non-stellar objects, as by definition, the majority of the flux of a
point source is contained within the point spread function.

In order to determine an empirical algorithm for classifying stellar
objects, we constructed a stellar object catalog from the
spectroscopic catalogs, \HST imaging, and these photometric data.
Eighty two objects were selected from the stacked \I band frame using
the IRAF task {\tt imexam} to measure the radial profiles of candidate
objects. Only those objects with both a Gaussian profile and a high
signal to noise peak flux were selected. Thirty additional objects
were selected from the CFRS spectroscopic data which had a reliable
redshift equal to zero, while an additional thirteen spectroscopic
stellar objects were selected from the DEEP data. Finally, we used the
\HST images to identify twenty seven additional sources,
providing a total of 152 stellar candidates to quantify the stellar
locus.

In addition, all objects which had a total I magnitude brighter than
twentieth magnitude were visually inspected and classified as stellar
or non-stellar. We constructed bounding boxes around the classified
stellar objects, providing empirical stellar classifications in each
band (\cf Figure~\ref{IStarClassify}). The final classification was
constructed by taking the union of the four separate classifications,
resulting in 505 stellar objects. The number-magnitude distribution of
stellar objects agrees with the predictions of the Bahcall-Soniera
model (1980). The spatial distribution of the stellar objects is
fairly random, with the possible minor exception of the image corners
where the PSF increases due to focal degradations.

\placefigure{IStarClassify}

\section{Deep Multi-band Optical Number Counts\label{numCounts}}

Counting the number of galaxies as a function of apparent magnitude is
a simple, yet powerful statistic in understanding the evolution of
galaxies. Measurements of the number-magnitude relation, however, have
typically been hindered in two ways. First, the amount of sky surveyed
was generally sacrificed in an effort to push the completeness limits
imaged deeper. This can produce biases due to object clustering and
sample variance. Second, most deep surveys do not fully sample the
optical spectral region. This limits the amount of evolutionary
information about the underlying galaxy population that can be
reliably extracted.

We, therefore, present the measured number-magnitude counts in the
optical bands \UBRI. This work is unique for the relatively large area
($\approx 196$ \SqArcminute) that has been deeply imaged in multiple
optical bands. As such, this dataset serves as a bridge between the
large area, shallow photographic surveys, and the small area, deep
fields such as the HDF (\cite{williams96}).

\subsection{Analysis}

Detecting faint objects in an image is hindered by many different
effects, including object superposition, noise spikes, as well as
redshift and evolutionary effects. Furthermore, these complications
strongly affect the number-magnitude relation. In order to reliably
extend the measured galaxy counts fainter, corrections for the
completeness of the survey and subsequent analysis must be
applied. Invariably, this requires heavy utilization of simulations:
either through the simulation of an entire image, or through the
addition of simulated images to the actual object frame. As we wanted
to determine the differential number-magnitude relation in a model
independent manner, we adopted the latter approach (\cf
Section~\ref{catalogLimits}).  Using the calculated completeness
function, we applied the appropriate scale factor to each galaxy as it
was added to the appropriate bin, rather than the entire bin
uniformly.  We estimated the errors in our sample by combining the
effects of Poisson noise with the one $\sigma$ completeness errors we
derived from our simulations.

We set the absolute lower limit used in our analysis to $\sim
10^{3.5}$ objects/Magnitude/\SqDegree, which corresponds to $\approx
100$ galaxies detected within a given half-magnitude bin over the
entire image, in order to limit the effects of poor statistics. We
measured the slope for the full range using a simple least squares
approach.  As the corrected counts of the bluer bands indicated a
change in the slope parameter ($\alpha$), we also measured the slopes
for the bright end (matching previous photographic surveys) and for
the faint end (which we can compare to the HDF). Modeling the
differential number magnitude counts as $dN(m) \propto m^{\alpha}$, we
measure a change in the slope ($\alpha_{B}$) in the $B_{AB}$ band
number magnitude relation at $B_{AB} \approx 24.4$ previously noted by
other authors (\cite{lilly91,metcalfe95}). We also measure a change in
the $U_{AB}$ band slope ($\alpha_{U}$) at $U_{AB} \approx 24.7$, which
has been suggested earlier (\cite{majewski89}). Interestingly, our
bright slopes agree quite well with the photographic data (\cf
Table~\ref{numCountTable}), and although the two normalizations
differ, the faint end slope of our ground based data also agrees with
our measurements of the slope of the corresponding HDF band
(\cite{metcalfe96}). The most likely explanation for the difference in
the faint end slopes lies in the differences in faint object detection
and extraction between the different surveys (\cf the different number
of objects detected in the HDF using different
techniques~\cite{ferguson98}.)

\placetable{numCountTable}

The measured number magnitude counts are presented in
Figure~\ref{uCounts} for the \U band, Figure~\ref{bCounts} for the \B
band, Figure~\ref{rCounts} for the \R band, and Figure~\ref{iCounts}
for the \I band. In each of these figures, previously published number
counts are also displayed. 

\placefigure{uCounts}
\placefigure{bCounts}
\placefigure{rCounts}
\placefigure{iCounts}

\section{Photometric Redshifts\label{photoZ}}

To understand the evolution of the universe, a large, uniform
spectroscopic sample must be utilized. The creation of such a sample
is extremely difficult and currently impractical. Many cosmological
tests, however, are more sensitive to the sample size (\ie Poisson
Noise) than small errors in distance --- which makes them perfect
candidates for utilizing a photometric redshift catalog. We have
developed an empirical photometric redshift technique
(\cite{connolly95,brunner97,myThesis}), which is not designed to
accurately predict the redshift for a given galaxy (\cite{baum62}) or
locate high redshift objects (\cite{steidel96}). Instead, it is
designed to provide distance indicators which are statistically
accurate for the entire sample, along with corresponding redshift
error estimates.

\subsection{Calibration Data}

The accuracy of any empirically derived relationship is predominantly
dependent on the quality of the data used in the analysis ---
photometric redshifts being no exception. As a result, we imposed
several restrictions on the calibrating data in order to minimize the
intrinsic dispersion within the photometric redshift relationship. The
calibration data was taken from the spectroscopic cross identification
with both the CFRS 14 hour +52 field and the DEEP \HST 5096 field (\cf
Section~\ref{spectroXID}).

The CFRS cross identification catalog consisted of 211 objects, while
the DEEP calibrating data provided 188 galaxies, with 28 duplicates
between the two. For all but two duplicate measurements, the mean of
the two redshifts was used. One of the remaining two duplicates (which
had quite discordant redshift measurements) was eliminated from the
sample due to source confusion, while the DEEP redshift was used for
the second (as it had a higher confidence) providing 370 spectral
cross-identifications. The next restriction was to select all objects
with $z > 0.0$ in order to remove the stellar objects, pruning the
catalog to 275 galaxies. 

Although the intrinsic error of a spectroscopic redshift is generally
quoted as $\delta < 0.001$, in reality however, a redshift is accurate
only when the spectral identification is also accurate. For the CFRS
data, a redshift quality was generated from both the spectral type and
the reliability of the redshift assigned (\cite{leFevre95}). The CFRS
galaxies were then restricted to the following six quality classes: 3,
4, 8, 93, 94, 98, to guarantee that only objects with redshifts having
a confidence greater than $95\%$ were retained; similar, albeit less
empirical, constraints were placed on the DEEP galaxies. The next step
was to restrict the sample to those objects which were below the
$10\%$ photometric error limit (\cf Section~\ref{catalogLimits} for
more information). In the remaining data, four objects were found to
have bad detection flags (\eg object near edge of frame, incomplete
aperture data) and were subsequently removed, while an additional four
galaxies were above our high redshift cut-off of $z = 1.2$. The final
spectroscopic calibration data consisted of 190 galaxies.

\subsection{The Empirical Relationship}

\subsubsection{Background}

The derivation of photometric redshifts from broadband photometry has
been previously shown to be more sensitive to broad spectroscopic
continuum features (primarily the break in the continuum spectra at
around 4000 \AA) rather than specific absorption/emission features
(\cite{connolly95}). As a result, we define five different redshift
intervals (\cf Table~\ref{algorithmZ}) which track the movement of the
4000 \AA \ break through our filter system with increasing redshift
(\cf Figure~\ref{effectiveFilters}). In the three intervals, low,
medium, and high, we can accurately approximate the galaxy
distribution in the four flux space \UBRI\ by a second order
polynomial. Between these selected redshift intervals, however, the
continuum break is moving between adjacent filters, introducing a
curvature in the galaxy distribution, necessitating the use of a third
order polynomial to accurately map the galaxy distribution.

\placetable{algorithmZ}

\subsubsection{Algorithm}
	
The 190 calibrating redshifts were, therefore, used to derive a global
third order polynomial in \UBRI\  which provided an initial redshift
estimate. Likewise, second order polynomials in \UBRI\  were determined
for the three different redshift intervals, while third order
polynomials were determined for the two break regions. The range of
calibrating redshifts for each polynomial fit was extended by
approximately 0.05 in order to diminish end-aliasing effects. This
algorithm is designed to generate an optimal redshift for objects by
using the more accurate local relations (\cite{brunner97}).  For each
derived polynomial fit, the degrees of freedom remained a substantial
fraction of the original data (a second order fit in four variables
requires 15 parameters while a third order fit in four variables
requires 35 parameters).

\subsection{Analysis of the Relationship}

The relative importance of the different bands in the individual
redshift intervals reflects the curvature inherent within the
distribution of galaxies in the four dimensional flux space. In a
given redshift range, the curvature can be accurately approximated by
a second order polynomial. Between redshift intervals, however, the
distribution displays a higher order curvature term (\cf the previous
discussion concerning the continuum break), which requires the higher
order fit. The correlation between the four band photometric and
spectroscopic redshifts is shown in Figure~\ref{fullPhotoZ}. The
intrinsic dispersion ($\delta z = 0.061$) is relatively stable
throughout the redshift range spanned by the calibrating galaxies, and
as shown in Figure~\ref{fullZHistogram}, is clearly Gaussian in
projection.

\placefigure{fullPhotoZ}
\placefigure{fullZHistogram}

The overall accuracy with which we can estimate redshifts leads us to
two related conclusions. First, this technique is extremely dependent
at these redshifts on the \hbox{$4000$ \AA\ } break over the redshift
interval sampled by the calibration data, which is present in nearly
all galaxies. Second, metallicity, dust, and age variations have
similar effects in this multidimensional space, albeit almost
orthogonal to the redshift vector (\cite{koo86}), and therefore affect
the galaxy distribution, which we are attempting to model, in similar
directions.

\subsubsection{Aperture Effects}

In a similar manner, we calculated photometric redshift relations for
the four different fixed aperture magnitudes in our catalog (2.5
pixels, 5 pixels, 10 pixels and 20 pixels). We compare their intrinsic
dispersions in Figure~\ref{apertureZ}. Due to the strength of the 4000
\AA\ continuum feature in determining the photometric redshift of an
object, we decided to test the hypothesis that ``bulge'' magnitudes
would produce a more accurate photometric redshift relationship. In no
case did a fixed aperture magnitude reduce the dispersion over a total
magnitude. The results, however, are encouraging, and worth a deeper
examination.

\placefigure{apertureZ}

The two best relations are clearly the 5 and 10 pixel aperture
magnitudes. In the $I$ band, the median full width at half maximum for
all objects with $I < 24.0$ is approximately 4.7 pixels. Thus, the 5
pixel aperture generally accounts for approximately $75 \%$ of the
object, while the 10 pixel aperture accounts for approximately $98 \%$
of the typical object. As the aperture size is decreased, the effects
of shot noise will become significant (since we are sampling only the
core of an object). On the other hand, as the aperture size is
increased, sky noise will begin to affect the relationship more
strongly (since we are sampling the wings of the object flux
distribution). Therefore, the total magnitudes that we use in our
analysis are best approximated by the 10 pixel fixed aperture
magnitudes, which sample the majority of an objects flux. Due to the
effects of ground based seeing, however, we can not reliably estimate
a bulge magnitude due to the faint nature of the galaxies in our
sample. This approach, however, might prove more useful for
space-based data (\ie \HST).

\subsubsection{Selection Effects}

A subtle, and often overlooked, effect in any photometric redshift
analysis is the requirement for accurate multi-band
photometry. Ideally we could restrict our photometric-redshift catalog
to only those objects which have measured magnitude errors below some
set limits (\eg $10\%$ photometry). This type of a restriction,
however, introduces two complications: a bias towards blue spectral
types, and a subsequently complicated selection effect (\cf
Figure~\ref{bandSelEffect}).

\placefigure{bandSelEffect}

In an attempt to overcome these biases, we restrict the full sample to
those objects which have both $I_{AB} < 24.0$ and measured magnitude
errors $< 0.25$ in \UBR. This minimizes any selection bias to only
faint early-type galaxies. In Figure~\ref{iBandCut}, the effects of
this cut can be discerned, and from the bottom panel (which models our
final selection criteria), it is fairly evident that this particular
selection results in a sample that is \I band limited with a high
redshift cut from the Lyman break systems (\ie $U$ and $B$ band
drop-outs). The remaining filter combinations contribute to the noise
in our analysis (\ie when we consider our final catalog complete to $I
\approx 24.0$), and amount to only a few percent when combined. The
final catalog contains 3612 sources, of which 442 are classified as
stellar objects, and a remaining 118 sources had bad detection flags
(\eg edge of frame, incomplete aperture data). Our final
photometric-redshift catalog, therefore, contains 3052 sources.

\placefigure{iBandCut}

\subsection{Error Analysis}

\subsubsection{Intrinsic Error Analysis}

By directly comparing the estimated photometric--redshift with the
measured spectroscopic redshift, we can estimate the precision with
which we have quantified the topology of the galaxy distribution in
the four band flux space \UBRI. When estimating redshifts for objects
with no spectroscopic redshifts (which is the goal of this analysis)
what is desired is an independent estimate of the error in the
photometric redshift.

In order to determine the optimal error estimate for galaxies with no
spectroscopic redshift, we developed a bootstrap error estimation
techniques that would generate multiple realizations of the
photometric redshift relation from which we could test various error
estimators. The fundamental principles behind this technique are to
model the effects of removing calibrating galaxies from the sample,
while simultaneously including the effects of the photometric
errors. This will simulate the possible effects of an incomplete
sampling of the distribution in flux space by the calibrating
galaxies.

Algorithmically, the calibration sample is randomly divided into two
samples, new calibration data, and test data. Although random, the
division is devised so that the calibrating sample is divided into 15
bins, with 14 of the bins ($\approx 177$ galaxies) used to calibrate
the relationship, and the remaining bin ($\approx 13$ galaxies) used
to test the new relationship. This algorithm was implemented in
Mathematica (\cite{wolfram96}), and 1000 iterations were produced. If
the division process was truly random, each galaxy would have a
probability of being selected of $1 / 15$ for each iteration, and,
therefore, in 1000 iterations, each galaxy should be selected on
average 67 times. The actual statistics compare quite well, as the
mean of the selection distribution was $\approx 65$, the standard
deviation was $\approx 20$, while the minimum number of times a galaxy
was selected was 43 and the maximum number was 185.

Every time a galaxy was a member of the test sample, its estimated
redshift was appended to the list of estimated redshifts for that
galaxy. After the completion of the 1000 iterations, the following
quantities were calculated for every calibration object from the list
of estimated redshifts: the mean of the estimated redshifts, the
trimmed (one sigma deviations from the actual mean) mean, the standard
deviation of the estimated redshifts, and the six quantiles (uniformly
spaced in $\sigma$) at the values: 0.0228 ($Q_{1}$), 0.1587 ($Q_{2}$),
0.3085 ($Q_{3}$), 0.6915 ($Q_{4}$), 0.8413 ($Q_{5}$), 0.9772
($Q_{6}$).

The six quantiles and the standard deviation of the distribution can
be used to define four independent error estimators. The advantage of
using the quantiles to estimate the standard deviation of a
distribution is that they are much less sensitive to extreme
outliers. Since the derivation of the photometric redshift
relationship is more dependent on certain calibrating redshifts (due
to the the incomplete sampling of the topology of the galaxy
distribution in the four dimensional flux space \UBRI\ by the
calibrating galaxies), the quantile error estimators are, on average,
more precise estimates of the redshift error than the standard
deviation of the distribution. After a lengthy comparison, we defined
the error of a photometric redshift to be $\sigma_{z} = (Q5 - Q2) /
2.0$. This was due in part to its tighter correlation with $\delta z$
than both the standard deviation and the $Q_{6 - 1}$ error estimator,
and also because the $Q_{5 - 2}$ error estimator samples a larger
number of data points than the $Q_{4 - 3}$ error estimator due to its
larger width (\cf Figure~\ref{sigZZ}).

\placefigure{sigZZ}

A comparison between the defined error estimator and the four
different magnitudes (\cf Figure~\ref{sigZM}) demonstrates the
sensitivity of the empirical fitting procedure on the calibration
sample. At both limits, the distribution of galaxies is not fully
sampled, and as a result, the estimated error is consequently
larger. The estimated error distribution is also affected by the
increase in photometric error at fainter magnitudes.

\placefigure{sigZM}

\subsubsection{Extrinsic Error Analysis}

To estimate the error in a photometric redshift for the full
photometric sample, we adopt a similar error estimation technique.
Different realizations of the photometric redshift relationship are
determined by adopting a bootstrap with replacement algorithm, in
which galaxies are randomly selected from the calibration sample and,
once selected, are not removed from the set of calibrating
galaxies. Thus, at the extremes, one galaxy could be selected 190
consecutive times or, alternatively, each redshift could be selected
exactly once (each of these realizations has the same
probability). This approach is designed to emphasize any
incompleteness in the sampling of the true distribution of galaxies in
the four dimensional space \UBRI\  by the calibration redshifts. As
before, with each different realization, the magnitudes of the
calibrating sample were drawn from a Gaussian probability distribution
function with mean given by the measured magnitude and sigma by the
magnitude error.

This algorithm was implemented in Mathematica, and 100 different
realizations of the photometric redshift relationship were
derived. For each different realization, a photometric redshift was
calculated for every galaxy in the photometric redshift catalog. The
six quantiles at the values: 0.0228 ($Q_{1}$), 0.1587 ($Q_{2}$),
0.3085 ($Q_{3}$), 0.6915 ($Q_{4}$), 0.8413 ($Q_{5}$), 0.9772
($Q_{6}$), were computed from the 100 different redshift estimates for
each object. The error in the photometric redshift for each object was
defined, as before, by $\sigma_{z} = (Q5 - Q2) / 2.0$.

The photometric redshift and corresponding error are compared in
Figure~\ref{fullSigmaZ}.  As expected the average estimated error is
the largest at the upper and lower redshift limits where the
incompleteness in the calibrating sample is most evident. The majority
of the rest of the objects with extremely large redshift errors are
blended in one or more bands (there are 361 objects with $\sigma_{Z} >
0.5$ and $86 \%$ of them are blended or contaminated by nearby
objects). As a result, objects which are affected by neighboring
objects are isolated from the high density surface delineated by the
majority of galaxies in the four flux space \UBRI . The affect these
objects impart on any subsequent analysis, however, is minimized by
the inclusion of their photometric error, which causes them to be
non-localized in redshift space. As a result, these objects provide a
minimal contribution to many ``redshift bins'' rather than strongly
biasing only a few bins.

\placefigure{fullSigmaZ}

On the other hand, those objects which are not blended, and still have
large estimated redshift errors could define an extremely interesting
sample for spectroscopic study (\ie QSO's or active galaxies). This
follows from the fact that by their very nature, these types of
objects are unlike the majority of galaxies in the universe.

\section{Spectral Classification\label{sedType}}

In addition to having an estimate for the redshift for each galaxy,
evolutionary trends can be discussed in terms of the different types
of observed galaxies present within the universe. For low redshifts,
galaxies are often segregated based upon their morphology (\ie the
Hubble sequence). At moderate to higher redshifts, however,
morphological typing is extremely difficult, especially for ground
based imaging. As a result, we adopt a different approach, which
classifies galaxies by the spectral type which best matches their
observed magnitudes.

\subsection{Synthetic Magnitudes}

The calculation of synthetic magnitudes from spectral energy
distributions (SEDs) is based on the definition of apparent
magnitudes.
\begin{equation}
m = -2.5 \log \left( \int f_{\lambda} R_{\beta}(\lambda) d\lambda \right) - m_{0}
\end{equation}
where $f_{\lambda}$ is the absolute spectral energy distribution of
the object in units of erg cm$^{-2}$ s$^{-1}$
\AA$^{-1}$. $R_{\beta}(\lambda)$ is the system response function for
filter $\beta \in \{$\U, \B, \R, \I$\}$ which we take as the
convolution of the CCD quantum efficiency and the appropriate filter
response function. $m_{0}$ is the zeropoint for the system, which is
generally derived for $\alpha Lyrae$ (Vega).  We transformed our
calculated synthetic magnitudes to the AB system (\cite{okeGunn83})
which introduces a zeropoint shift $\Delta_{AB}$ given by computing
the synthetic AB magnitudes for $\alpha Lyrae$ in each filter (\cite{frei94}): $\Delta_{AB}(\U) = 0.730$, $\Delta_{AB}(\B) = -0.055$,
$\Delta_{AB}(\R) = 0.187$, and $\Delta_{AB}(\I) = 0.439$.

We, therefore, calculated magnitudes in a filter $\beta$ given a
spectral energy distribution,
\[
m^{\beta}_{AB} = -2.5 \log \left( \int f_{\lambda} R_{\beta}(\lambda) d\lambda
\right) - m^{\beta}_{0} + \Delta_{AB}
\]
where the zeropoints are $m^{U}_{0} = 15.20$, $m^{B}_{0} = 13.89$,
$m^{R}_{0} = 14.17$, and $m^{I}_{0} = 15.16$.

\subsection{Template Spectral Energy Distributions}

In order to match the full range of observed magnitudes in the
photometric redshift catalog, we required template spectra which
extend to $\lambda = 3100$ \AA$/2.2 \approx 1400$ \AA, where $3100$
\AA\ is the approximate blue cut-off for the \U filter. Rather than
generating model template spectra to classify galaxies, we selected
the Coleman, Wu, and Weedman (1980) UV-optical template spectra
constructed from real galaxy spectra. The sample includes Irregular,
Scd, Sbc, and Elliptical spectral energy distributions. The four
template spectra are compared in Figure~\ref{spectraNorm} in which the
spectra smoothed over $10$\AA, and normalized to the Elliptical flux
at $\lambda = 5500$\AA.

\placefigure{spectraNorm}

Synthetic AB magnitudes for the four template spectra were calculated
for redshifts from $z = 0.0$ to $z = 1.2$ in steps of $\Delta z =
0.05$. This involves both redshifting each template spectra so that
$\lambda^{\prime} = \lambda/(1 + z)$, and reducing the flux by a
factor of $(1 + z)$.

\subsection{Spectral Classification Algorithm}

To determine a spectral classification for each object, we need to
optimally select a spectral energy distribution from the grid of
redshifted template spectra. Algorithmically, we need to select the
best fit to our data for different models, which is most easily done
using chi-square fitting (\cite{press92}). The optimal spectral energy
distribution will, therefore, minimize the chi-squared statistic:
\[
\chi^2 = \sum_{\beta}^{4} \left( f_{\nu}^{\beta} - \gamma
t_{\nu}^{\beta} \right)^2
\]
where $\beta \in \{$\U, \B, \R, \I$\}$, $f_{\nu}^{\beta}$ is the flux
for the target object in the $\beta$ filter, and $t_{\nu}^{\beta}$ is
the flux for the current template spectra in the $\beta$ filter. The
constant $\gamma$ is determined by minimizing the variation of
$\chi^2$ with respect to $\gamma$, which gives:
\[
\gamma = \frac{\sum_{\beta}(f_{\nu}^{\beta}
t_{\nu}^{\beta})}{\sum_{\beta}(t_{\nu}^{\beta})^{2}}
\]

This classification technique is essentially an inverted template SED
photometric redshift calculation.  The quantities $\gamma$, and
$\chi^{2}$ are computed for each of the redshift-selected template
spectra from the object's observed broadband AB magnitudes and
magnitude errors (\ie to include the effects of photometric
errors). This process was extended to include the estimated redshift
errors, by selecting the redshift of a particular object from a
Gaussian probability distribution function with mean and sigma given
by the object's estimated redshift and redshift error. This was
performed one hundred times for each object, resulting in one hundred
different spectral classifications. The particular template which had
the smallest $\chi^2$ was then selected as the optimal SED for that
object.

The different percentages of each spectral type are compared in
Figure~\ref{sedPercentage}. Clearly the bluer spectra dominate the
classification. This should not be interpreted as evidence for an
overwhelming dominance of ultra luminous galaxies in this sample,
particularly in light of our bias against early type
galaxies. Instead, the best interpretation is that from the four
template spectra originally selected for this analysis, the redshifted
SEDs of nearby starbursting galaxies best match the observed data. In
fact, this result demonstrates that galaxies tend to become bluer with
redshift (\ie the faint blue galaxy problem). Other examples of this
are seen in the CFRS luminosity functions (\cite{lilly95}) and the
HDF morphological number counts (\cite{driver98}).

\placefigure{sedPercentage}

To form an impression of the accuracy of this spectral classification,
each template was compared to the objects which were assigned to that
template in two different colors: $U_{AB} - I_{AB}$
(Figure~\ref{tui}), which presents the largest spectral baseline; and
$(B_{AB} + R_{AB}) / 2.0 - I_{AB}$ (Figure~\ref{tbi}), which is an
interpolated $V_{AB}$ band magnitude.  For all four templates, the
data clearly agree quite well, indicating that the classification
algorithm is working properly.

\placefigure{tui}
\placefigure{tbi}

\section{Application: $N(z_{P})$\label{nOfZApp}}

We can derive the number of galaxies as a function of redshift,
$N(z_{P})$, as an example of a statistical application of photometric
redshifts. First, we computed the number redshift distribution (in the
traditional fashion) for redshift bins of width 0.1 magnitude. The
resulting histogram is compared with the measured number-redshift
distribution for the combined DEEP and CFRS samples in
Figure~\ref{nOfZ}. Since photometric redshifts are not equivalent
to spectroscopic redshifts, this is clearly not the optimal method.

\placefigure{nOfZ}

An analytic method for estimating the number redshift distribution
using photometric redshifts can be used to provide a more realistic
redshift distribution. We define the probability density function,
$P(z)$, for an individual galaxy's redshift to be a Gaussian
probability distribution function with mean ($\mu$) given by the
estimated photometric redshift and standard deviation ($\sigma$)
defined by the estimated error in the photometric redshift.

\[
P(z) = \frac{1}{\sigma \sqrt{2\pi}} e^{\left(-\frac{(z -
\mu)^2}{2\sigma^2}\right)}
\]

When constructing the number redshift distribution, the cumulative
distribution function for each galaxy is calculated over each redshift
bin. This requires numerically integrating the probability density
function, which is given by the Error function, $erf(z_{P})$, for each
galaxy between the endpoints of each redshift bin. 

A formal, analytic technique, however, is not always available to
utilize photometric redshifts and their associated errors when
measuring cosmologically interesting quantities. As a result, we have
developed an alternative technique, the galaxy ensemble
approach. Essentially, we treat the problem in the context of
statistical mechanics, where each galaxy is localized in redshift
space by a Gaussian probability distribution function. To calculate a
physically meaningful quantity, we create multiple realizations (or
ensembles) of the galaxy redshift distribution, and calculate the
appropriate quantity for all of the different ensembles. We then
average the different measurements to produce the desired value,
simultaneously producing a realistic error estimate.

To demonstrate the viability of this technique, we calculated the
number redshift distribution as a function of spectral type, both
analytically and using the ensemble approach. From
Figure~\ref{nOfZTypes}, the two distributions show remarkable
agreement, both with each other as well as the spectroscopic 
number redshift distribution, with the benefit of the ensemble error estimate
demonstrated (\ie error bars).  The differences in the redshift
distributions of the different spectral types, which were computed as
outlined in Section~\ref{nOfZApp}, such as the selection bias against
Elliptical galaxies, are also visible.

\placefigure{nOfZTypes}

\section{Conclusion}

In this paper we have presented the statistical technique to quantify
galaxy evolution. We also presented the photometric data which we have
used to develop this technique. Using the number magnitude test, we
have verified the validity of our catalog, and demonstrated the
turnover in the $U$ and $B$ band number counts previously
discussed. The empirical photometric redshift relation we derived has
an intrinsic dispersion of $\delta z = 0.061$ out to $z = 1.2$, and is
Gaussian in projection.

In order to more realistically quantify the evolution of galaxies, we
also developed techniques to estimate the error in a photometric
redshift, and classify the catalog objects by spectral type. To
demonstrate the effectiveness of this approach, we presented the
number redshift distribution as a function of spectral type using two
different techniques: the analytic approach, and the ensemble
approach. In the future, we plan on using the ensemble approach to
measure the evolution of the luminosity function and the angular
correlation function with redshift and spectral type
(\cite{myThesis,connolly98}).

\acknowledgments

First we wish to acknowledge Gyula Szokoly for assistance in obtaining
the data. We also would like to thank Barry Lasker, Gretchen Greene,
and Brian McLean for allowing us access to an early version of the GSC
II. We also wish to acknowledge useful discussions with Mark
Dickinson, George Djorgovski, Mark Subbarao, and David Koo. We would
also like to thank the referee, Rich Kron, for valuable suggestions on
improving this work. RJB would like to acknowledge support from the
National Aeronautics and Space Administration Graduate Student
Researchers Program. AJC acknowledges partial support from NASA grant
AR-06394.01-95A. ASZ has been supported by the NASA LTSA program.

\newpage

\figcaption[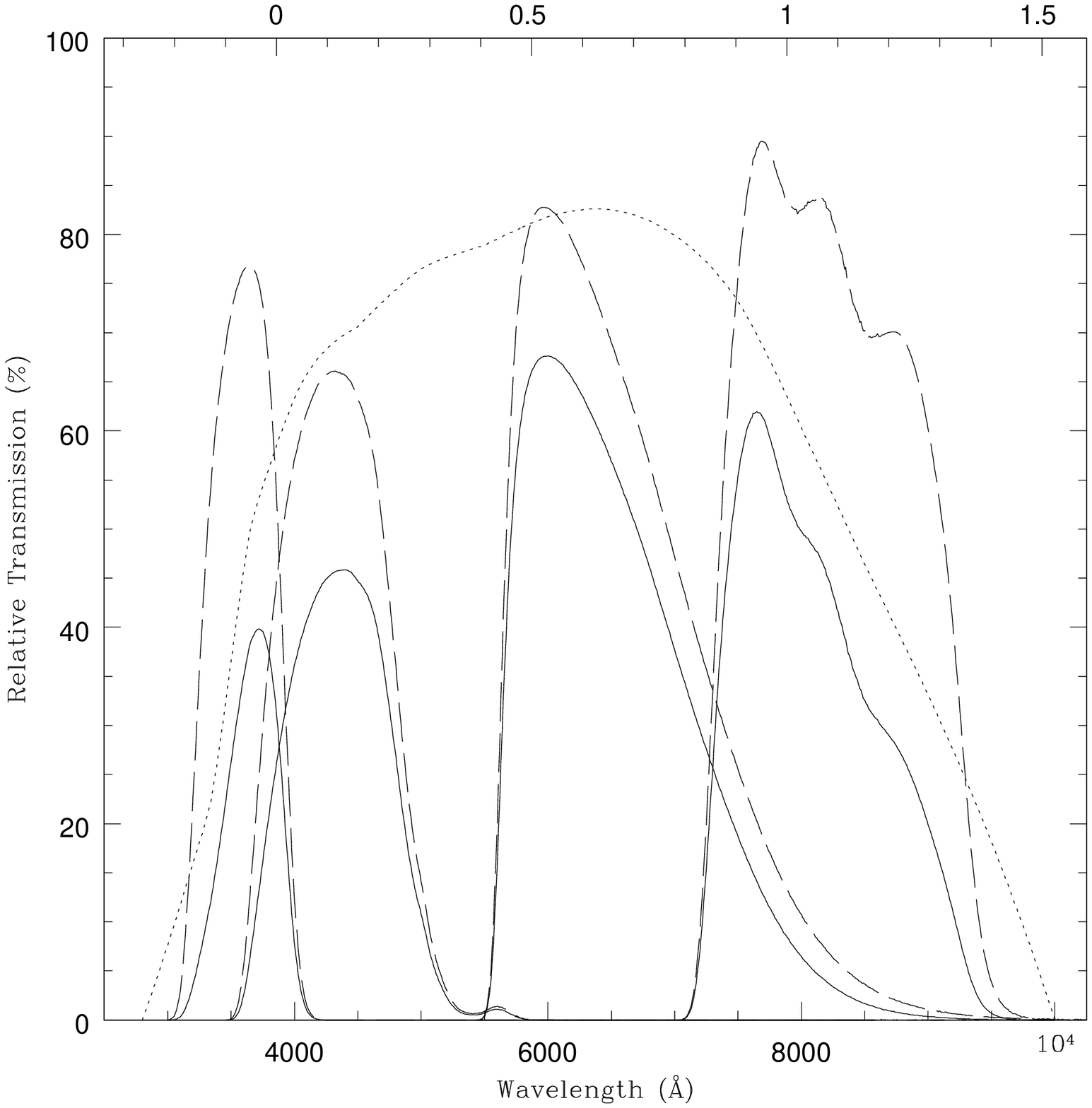]{The effective transmission for the 
standard broadband filters \UBRI\ available for use at the KPNO 4
meter. The dashed lines are the individual filter transmission curves,
the dotted line is the CCD detection quantum efficiency curve, and the
solid curves are the convolution of the filter and detector quantum
efficiency. Across the top is displayed the equivalent redshift of the
4000 \AA\ break. The data points were obtained via anonymous FTP from
the KPNO archives.\label{effectiveFilters}}

\figcaption[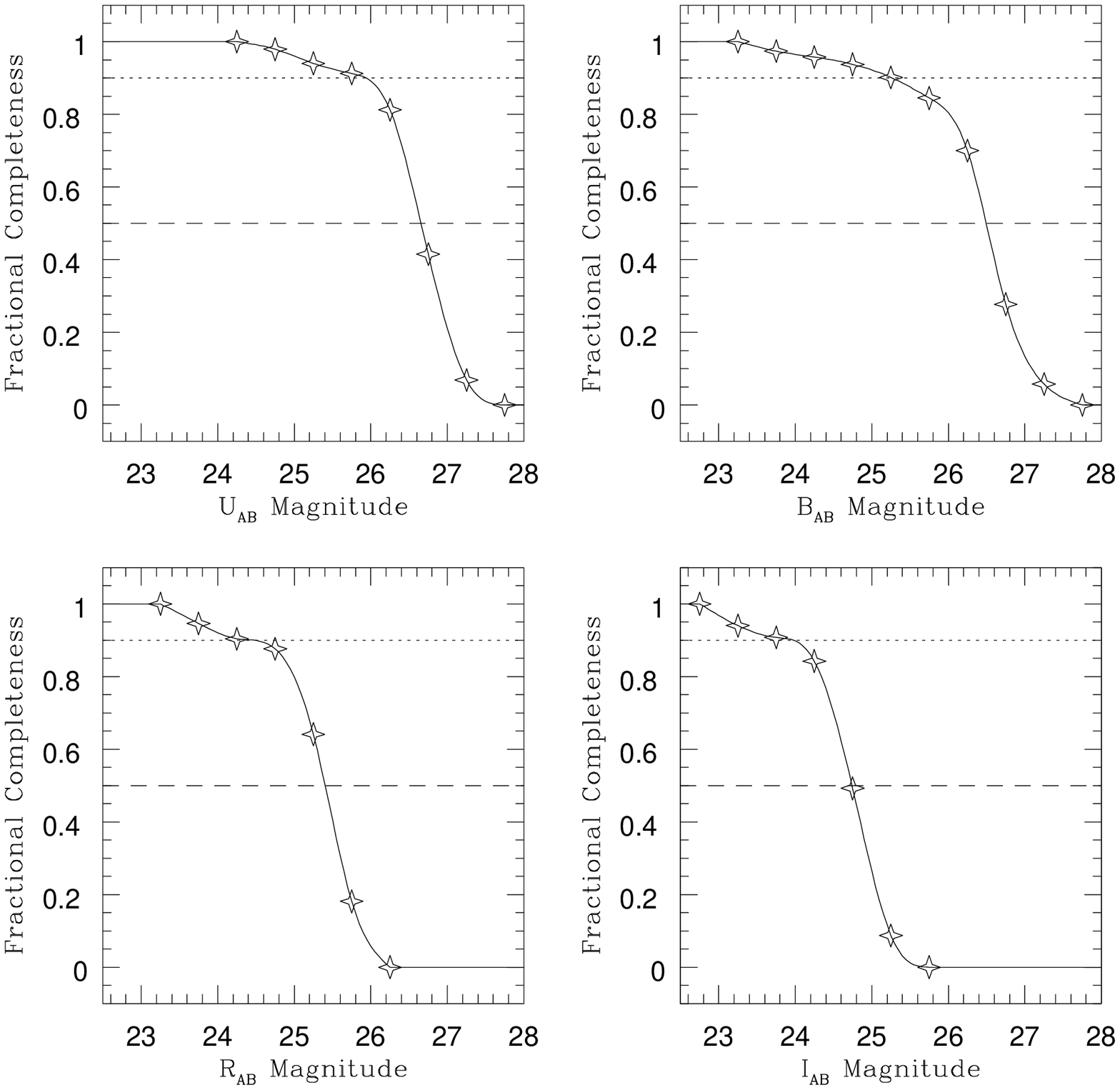]{The fractional completeness curves for 
all four bands. The short dashed line in each figure is the $90\%$
completeness limit, and the long dashed line is the $50\%$
completeness limit. The starred points are the values derived from the
simulations. The curves are the second order interpolating polynomial
fits to the points determined from simulations.\label{completeness}}

\figcaption[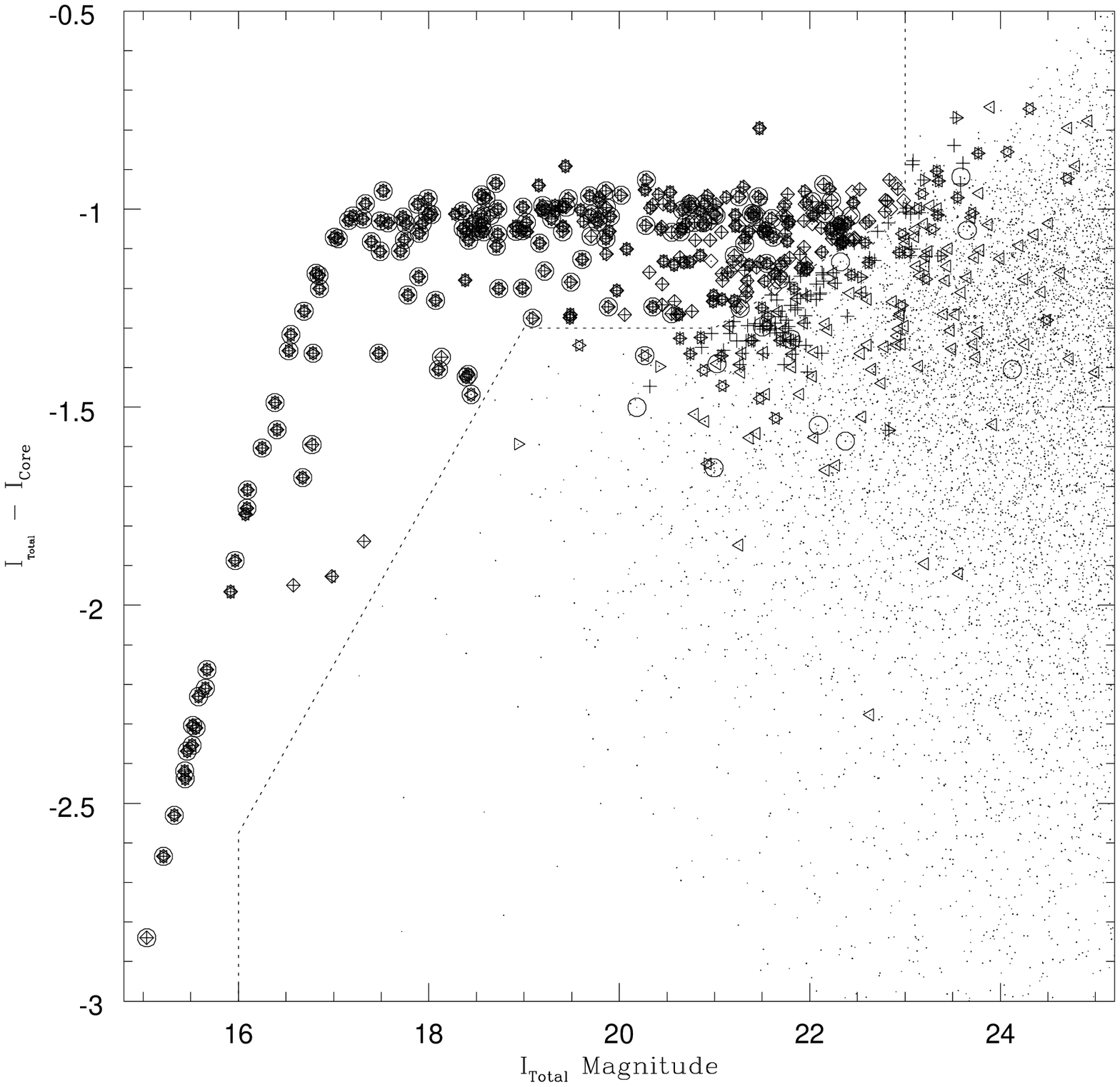]{The I Band stellar classification. The 
small dots are all of the galaxies in the catalog. The dotted line
delineates the stellar classification criteria. The large circles
indicate objects which are from the classification training set. The
remaining symbols: diamond, plus, left triangle, and right triangle,
indicate objects that were classified as stars in the \I, \R, \B, \&
\U\ bands, respectively.\label{IStarClassify}}

\figcaption[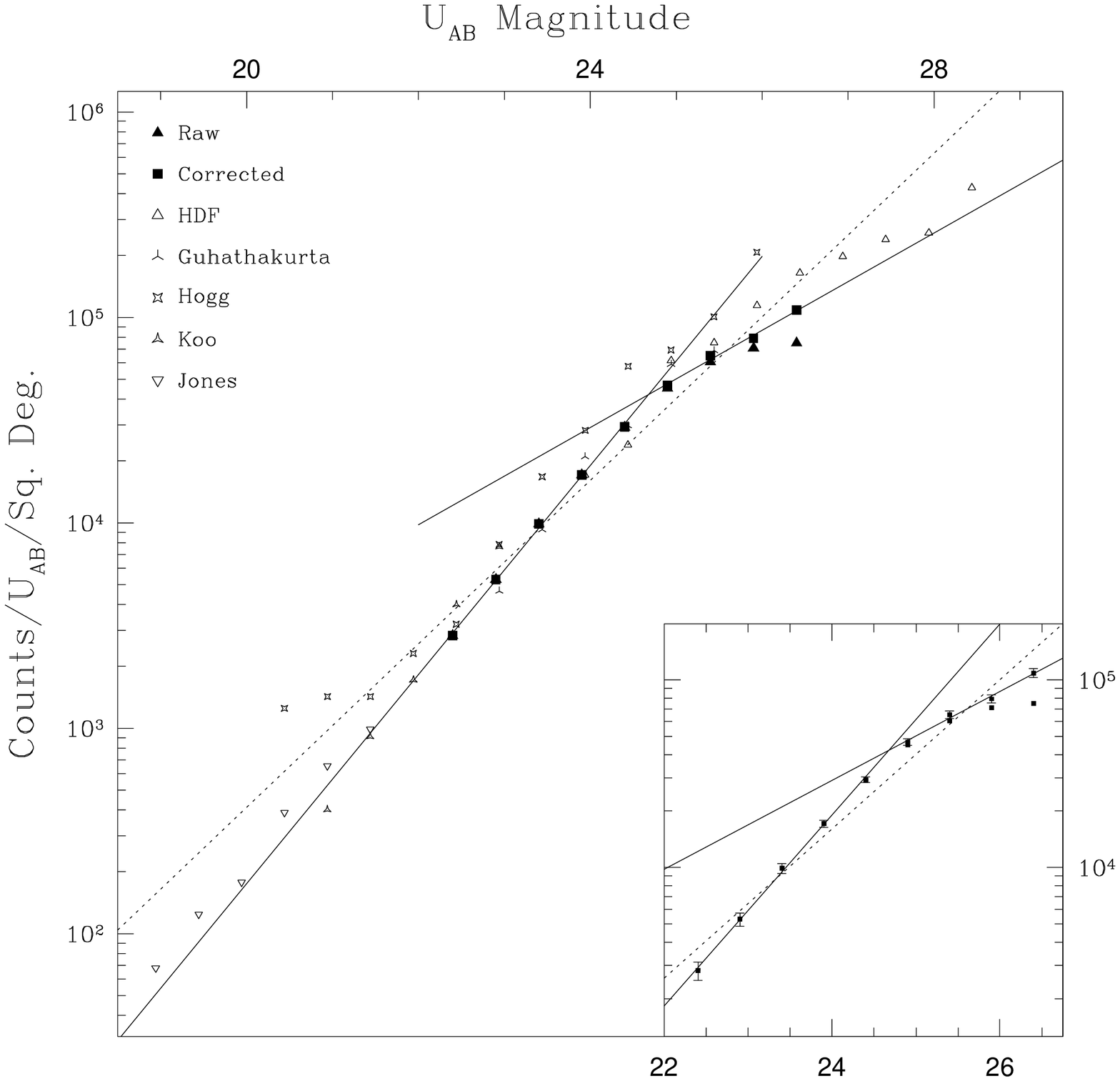]{The U Band number counts. Also plotted are 
comparable published counts, and the global fit (dashed line) and
high/low fits (solid lines). The inset in the lower left corner
indicates the errors and fits in more detail.\label{uCounts}}

\figcaption[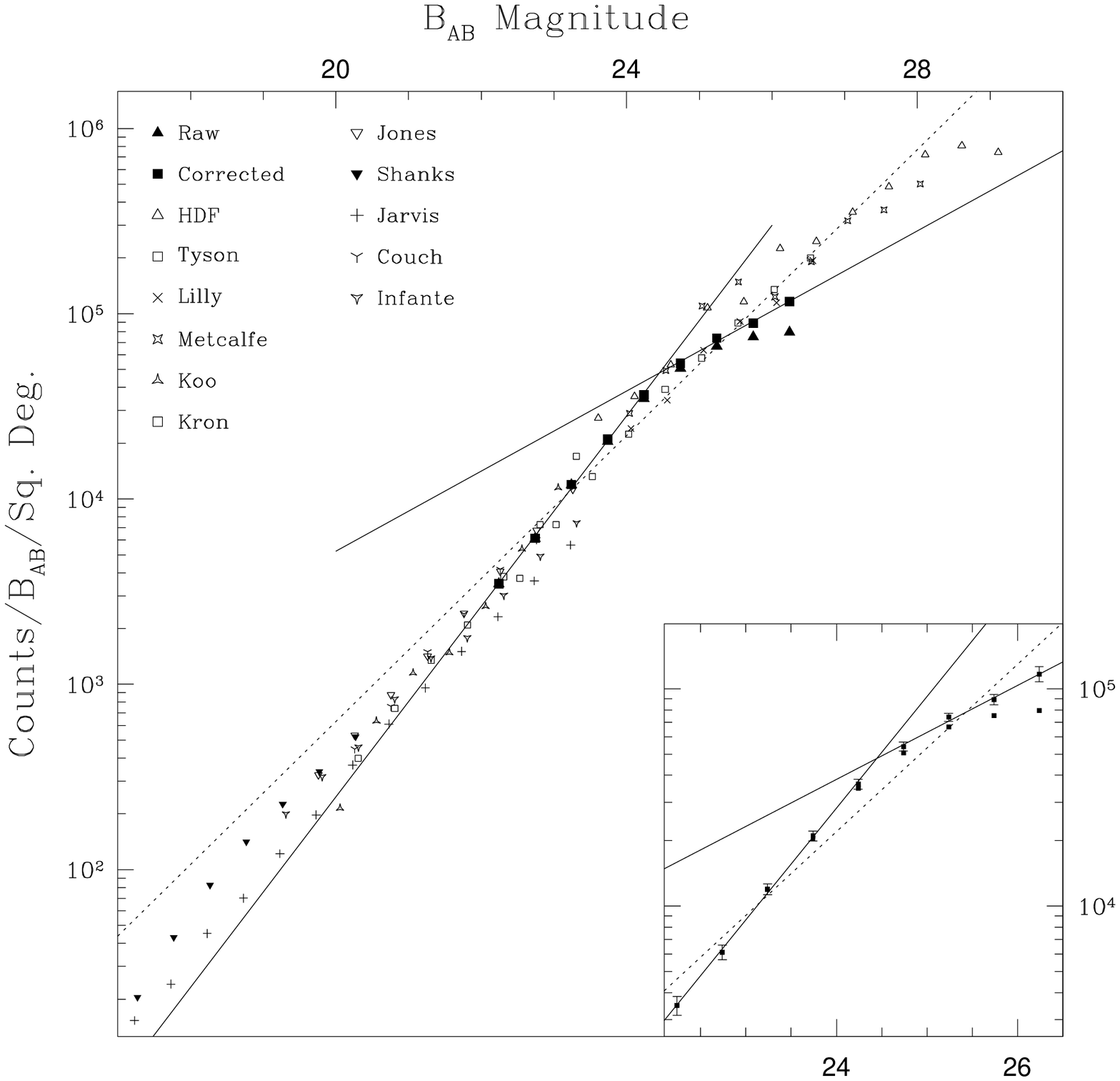]{The B Band number counts. Also plotted are 
comparable published counts, and the global fit (dashed line) and
high/low fits (solid lines). The inset in the lower left corner
indicates the errors and fits in more detail.\label{bCounts}}

\figcaption[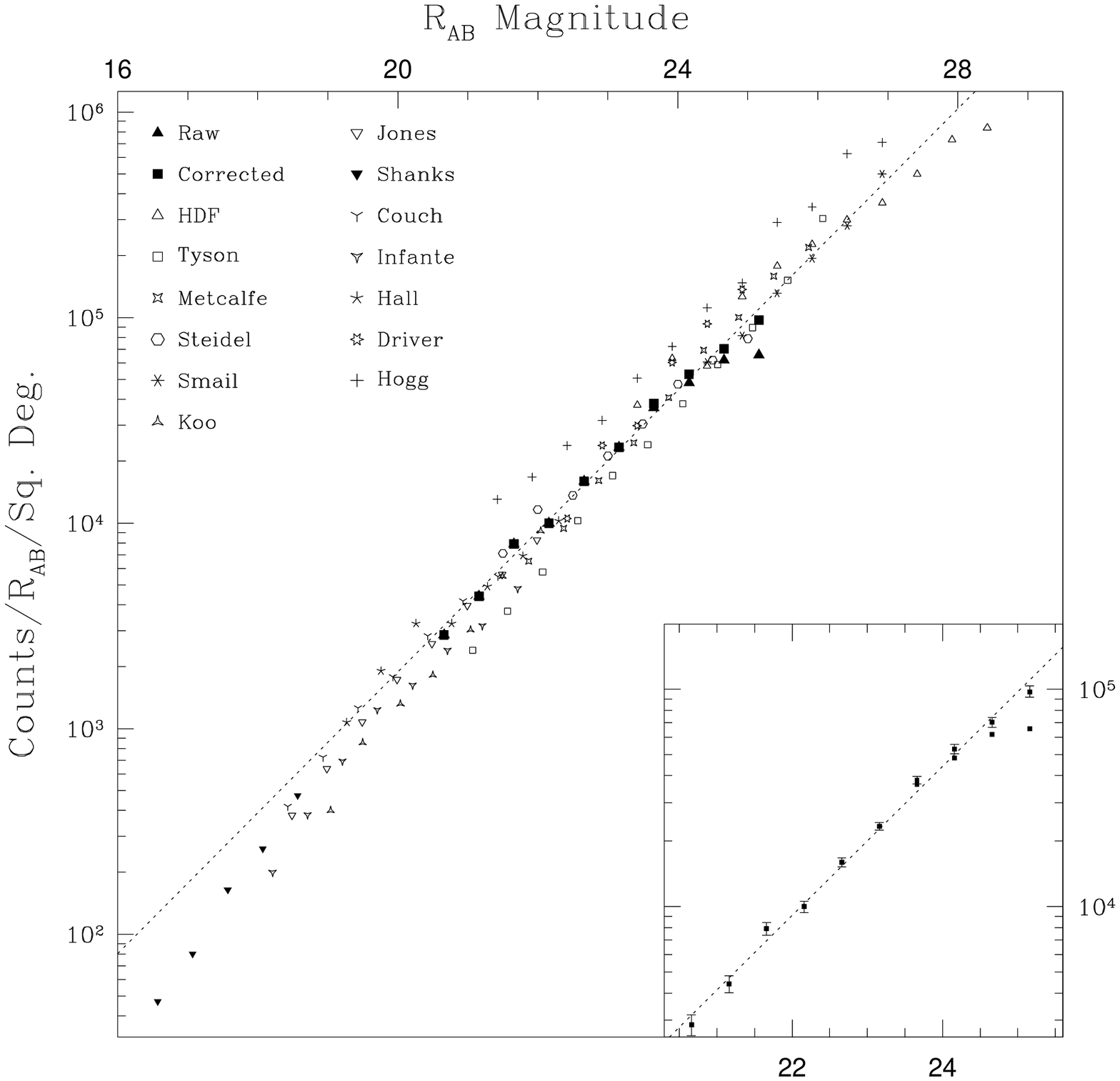]{The R Band number counts. Also plotted are 
comparable published counts, and the global fit (dashed line). The
inset in the lower left corner indicates the errors and fits in more
detail.\label{rCounts}}

\figcaption[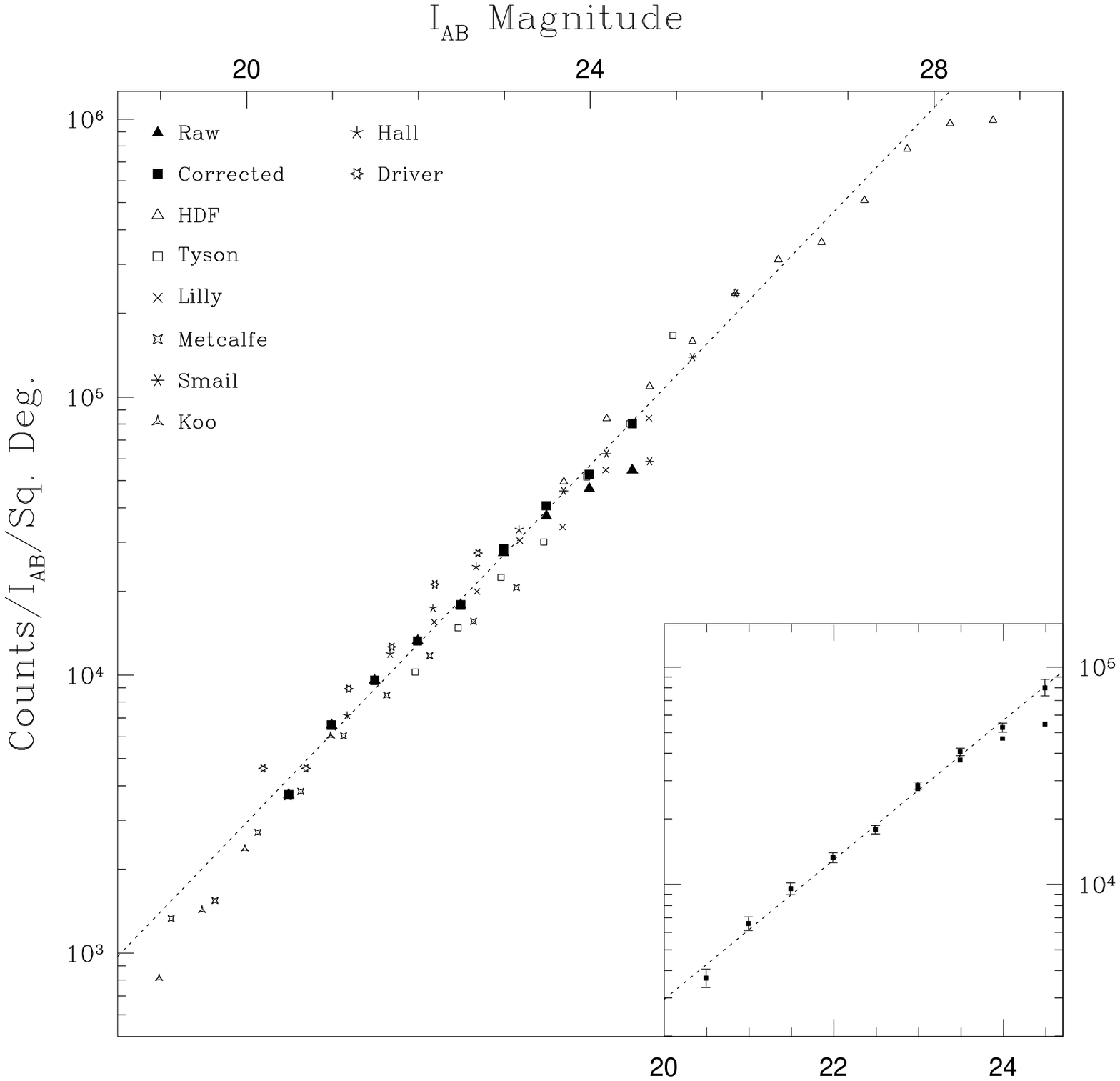]{The I Band number counts. Also plotted are 
comparable published counts, and the global fit (dashed line). The
inset in the lower left corner indicates the errors and fits in more
detail. \label{iCounts}}

\figcaption[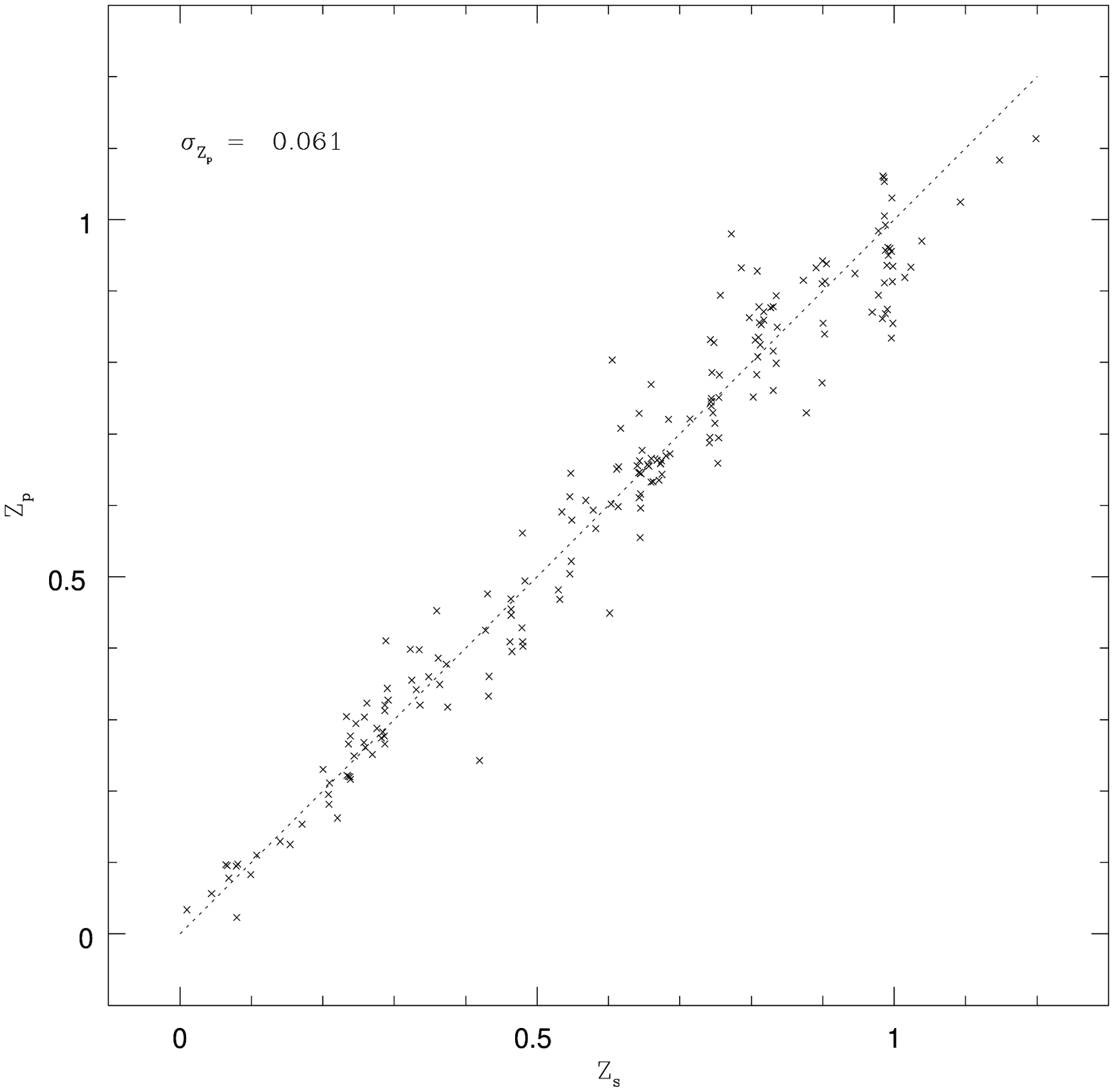]{Correlation between photometric and 
spectroscopic redshifts for the entire calibration sample. The
straight line is of unit slope, and is not a fit to the actual
data.\label{fullPhotoZ}}

\figcaption[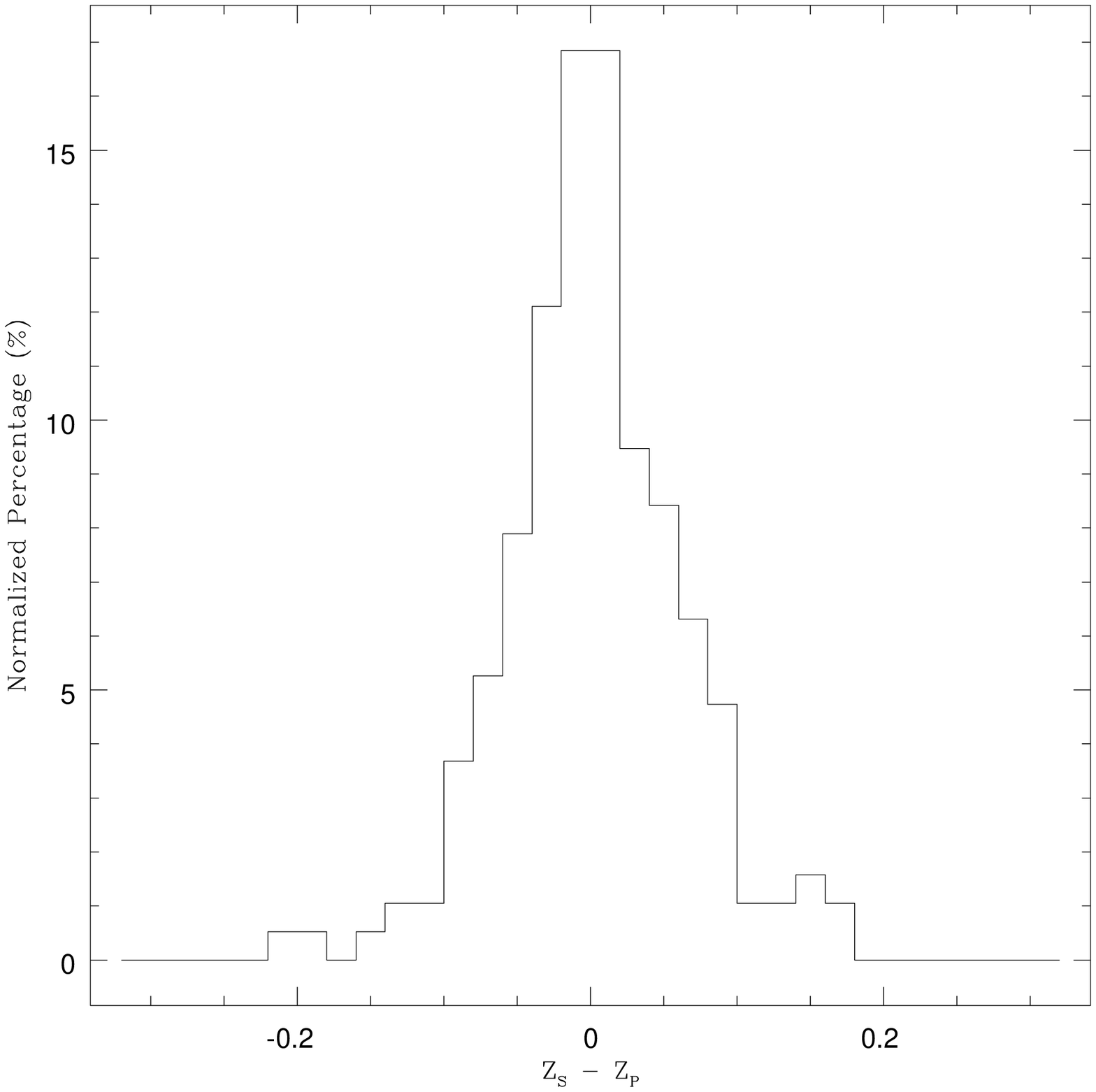]{A histogram of the residual 
differences between photometric and spectroscopic redshifts for the
entire calibration sample.\label{fullZHistogram}}

\figcaption[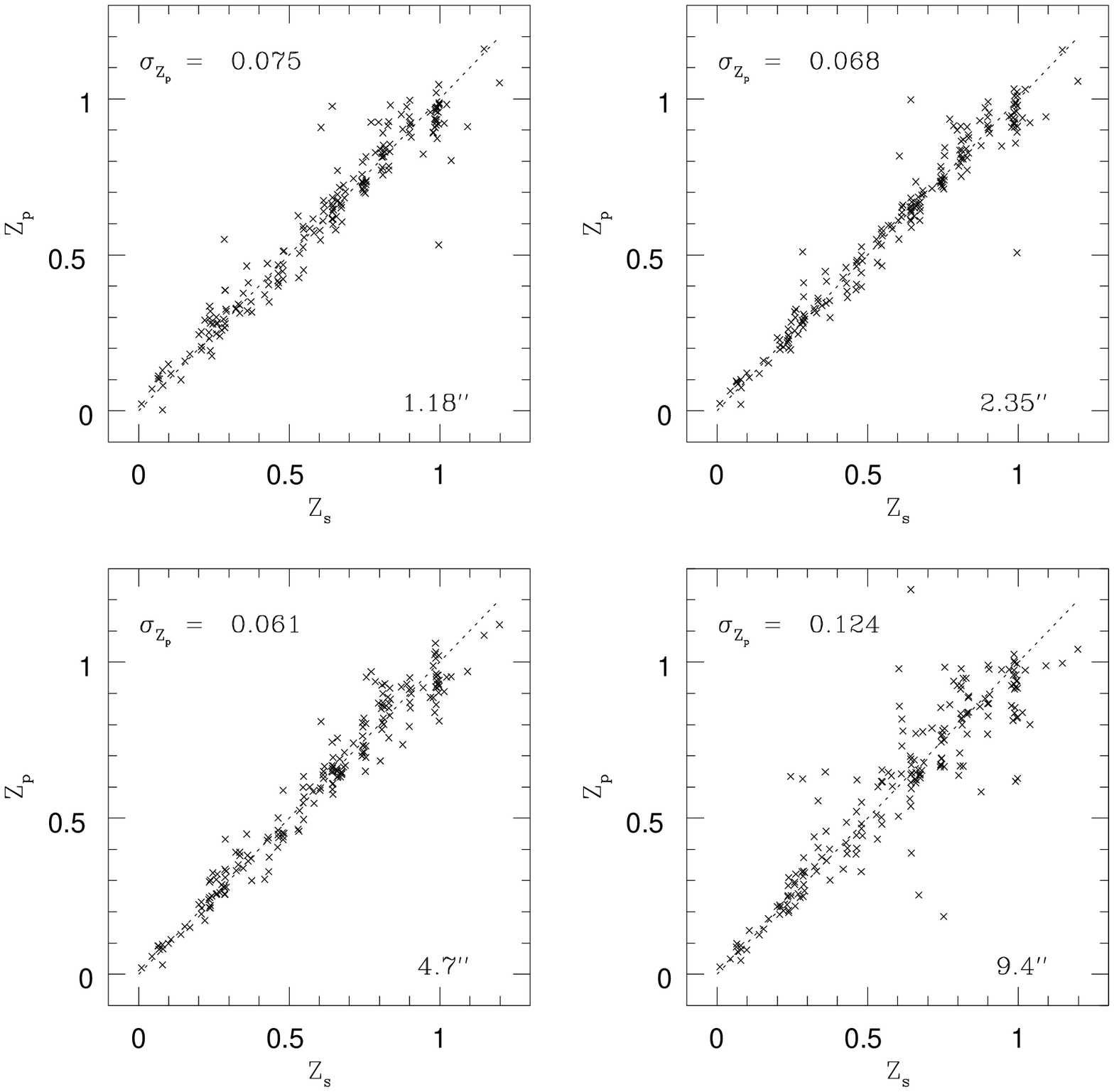]{A comparison between the aperture 
magnitude photometric redshifts and the calibrating spectroscopic
redshift sample.\label{apertureZ}}

\figcaption[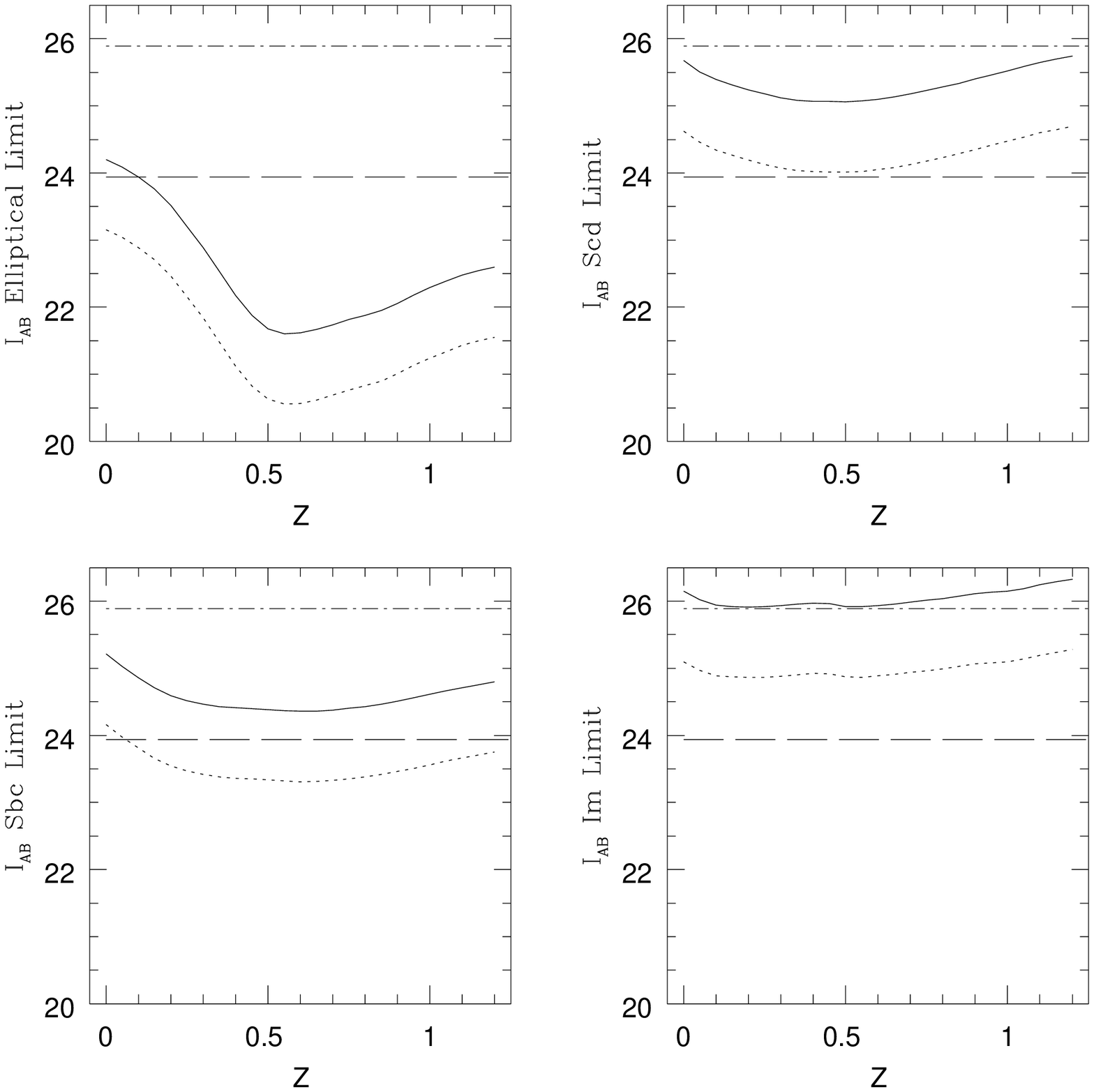]{The selection effect that results from 
requiring accurate four band photometry.  In each figure, the dash-dot
line is the \U band $10\%$ photometric limit, while the long dash line
is the \I band $10\%$ photometric limit. The two curves trace the $U -
I$ color for a source of a given SED type that has a \U band magnitude
at the $10\%$ photometric limit (dotted line) and $25\%$ photometric
limit (solid line). The exclusion of the early types at moderate to
large redshifts is clearly present. The other types, however, are
clearly not excluded at any redshift in our final
catalog.\label{bandSelEffect}}

\figcaption[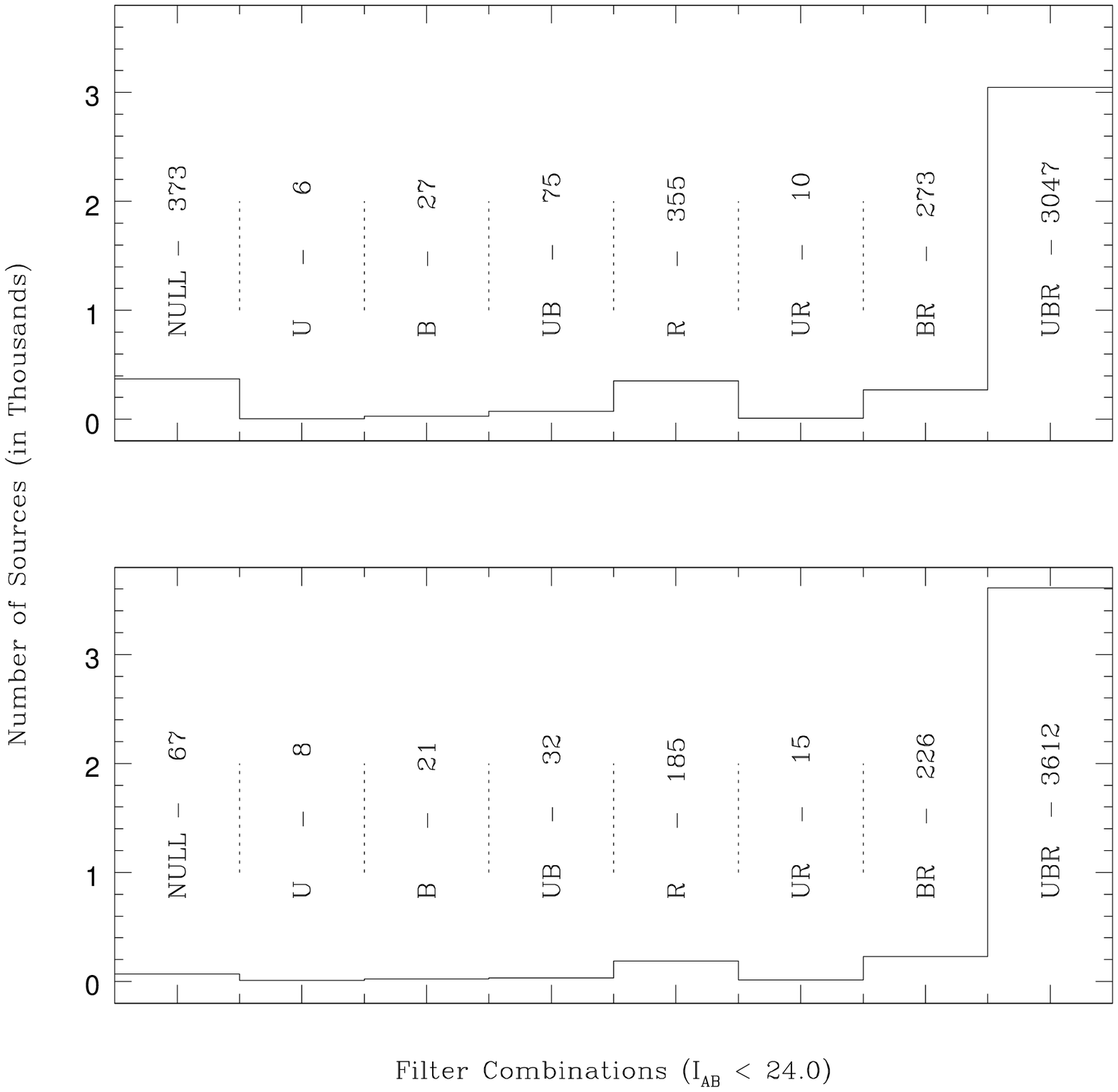]{The band distribution of sources with 
$I_{AB} < 24.0$ and $10\%$ photometric errors (top panel) and $25\%$
photometric errors (bottom panel). The horizontal axis indicates the
photometric errors in the other three bands (\ubr). The first bin
contains the objects which have magnitude errors greater than the
appropriate limit (or were not even detected in a band). Note the two
secondary peaks in the bottom panel at {\em R} and {\em BR}. A closer
inspection reveals that a majority of these sources are most likely \U
and \B drop-out systems (Steidel \etal 1996). The small size of the
other bins demonstrates their contribution to the statistical noise in
our analysis.\label{iBandCut}}

\figcaption[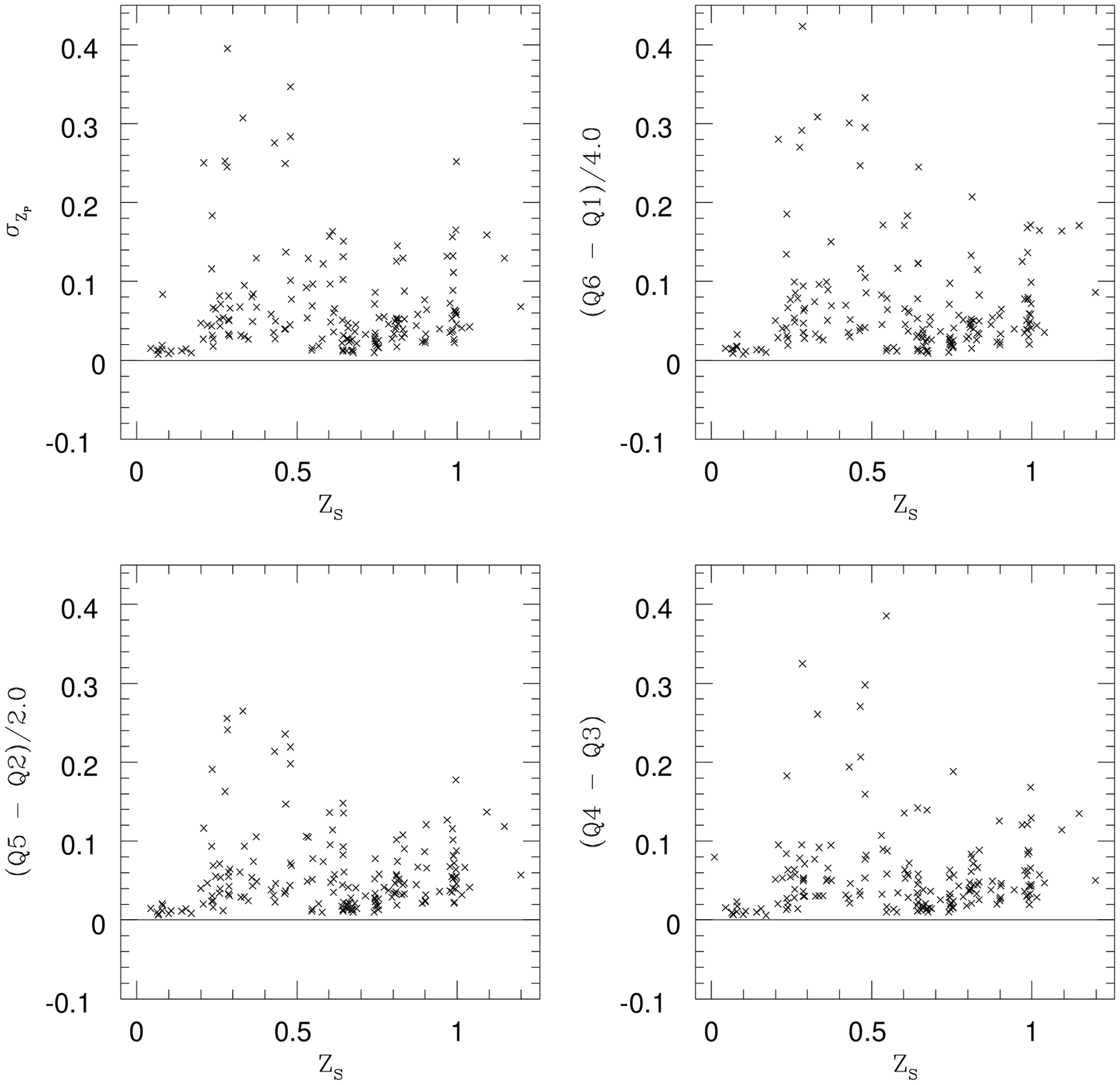]{A comparison between the different error 
estimates and the spectroscopic redshift.  The figure has been trimmed
for clarity to display only those sources with errors less than
0.45. This eliminated 14 objects from the top left figure, 12 objects
from the top right figure, 10 from the bottom left figure and 6 from
the bottom right figure. The estimated error can be quite large where
the number of calibrating galaxies is insufficient to model the galaxy
distribution in flux space.The $Q_{5 - 2}$ error estimate (bottom
left) has the smallest scatter.\label{sigZZ}}

\figcaption[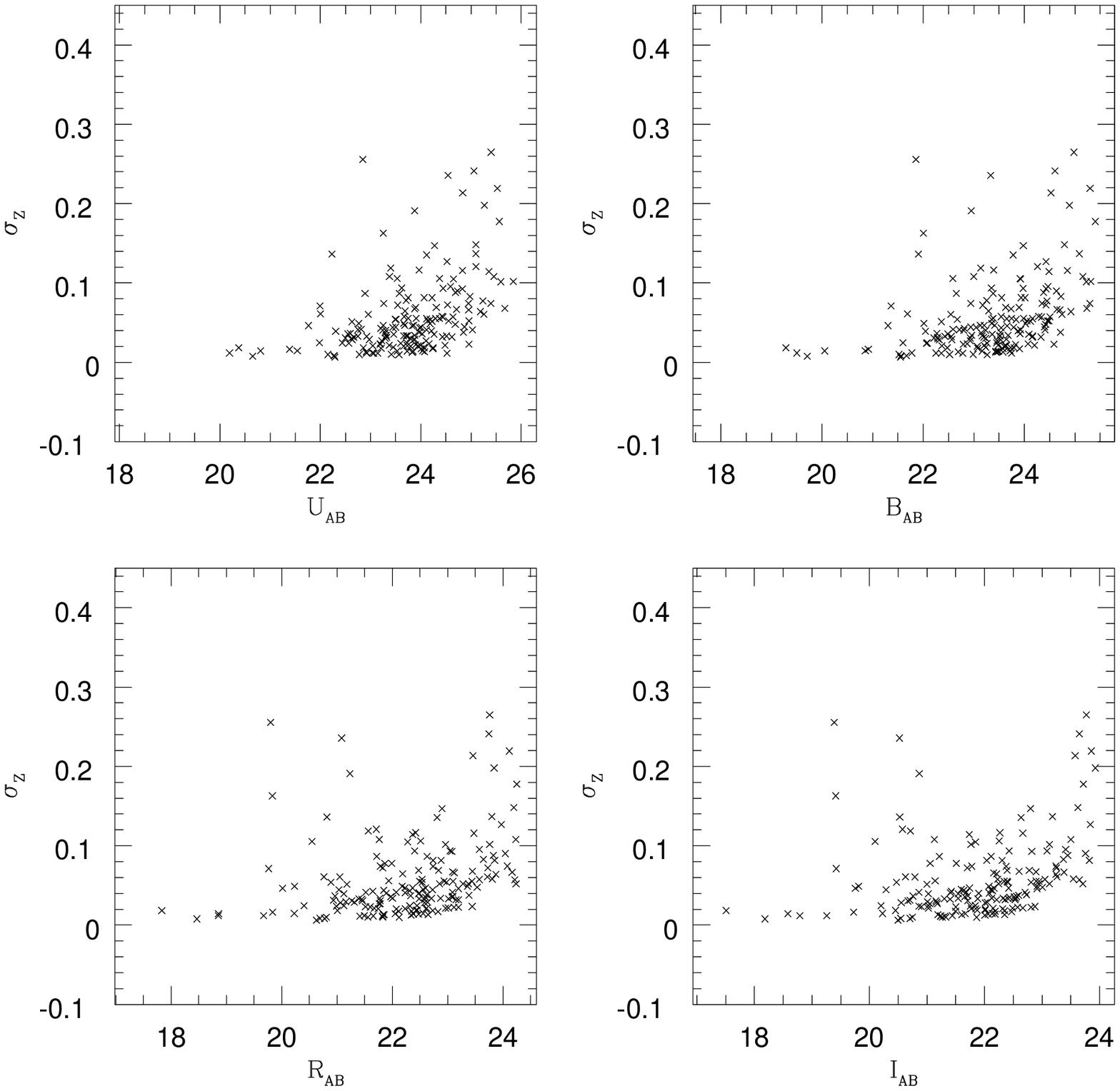]{A comparison between the photometric 
redshift error estimator and the magnitudes of the calibrating
galaxies. The figures have been trimmed for clarity, removing 10
calibrators.\label{sigZM}}

\figcaption[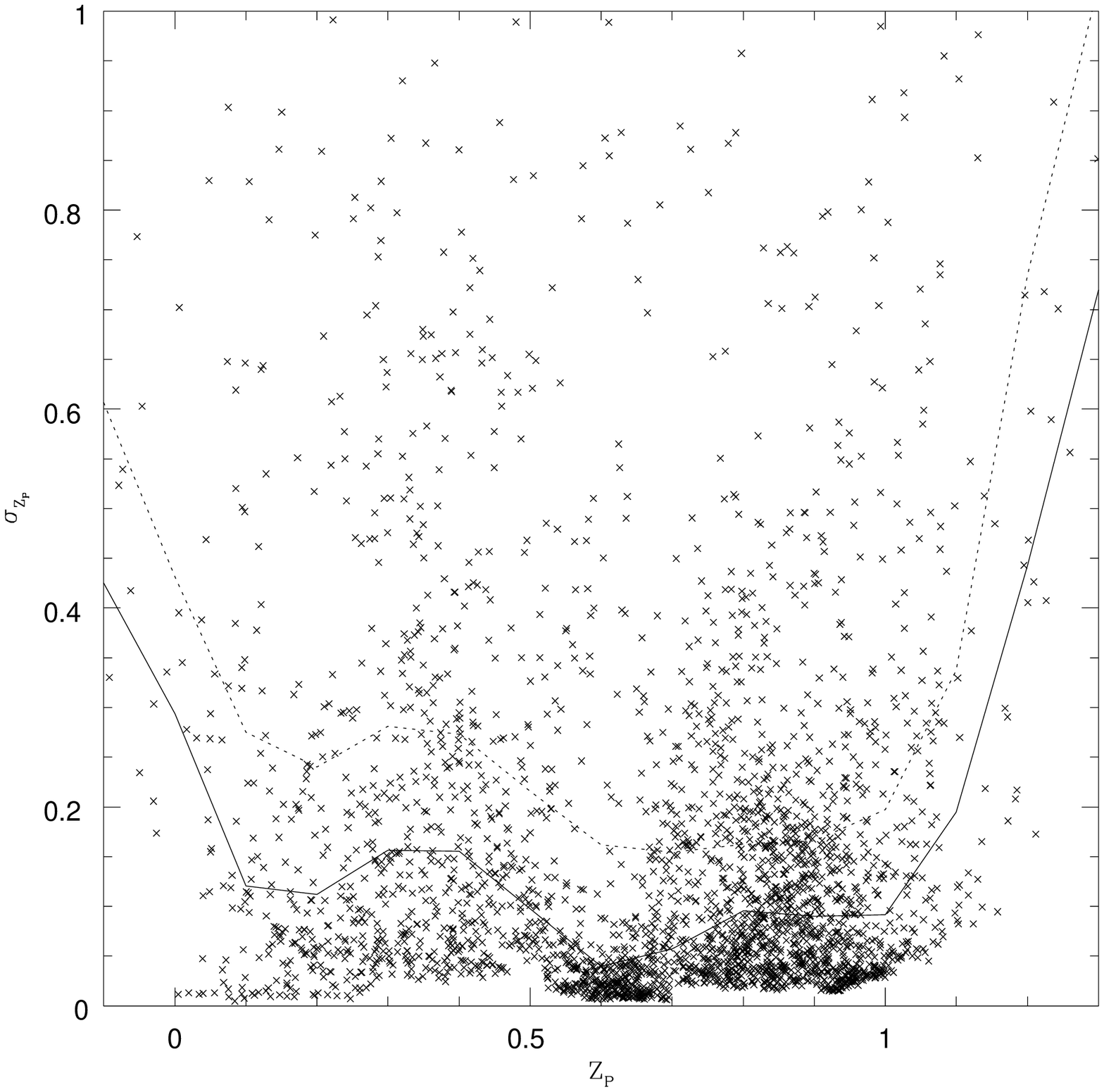]{Correlation between photometric redshifts 
and the estimated redshift error for the significant catalog
objects. Of the total sample of 3052 sources, 113 have photometric
redshifts outside the range $z \in [0.0, 1.2]$. Also plotted are the
median (solid line) and mean (dotted line) of the estimated
photometric redshift errors, binned by 0.2 in redshift. The increase
in both the mean and median of the binned estimated redshift error,
reflect the built-in error correction, as these objects are
down-weighted in any analysis since their redshift probability
distribution functions are non-localized.\label{fullSigmaZ}}

\figcaption[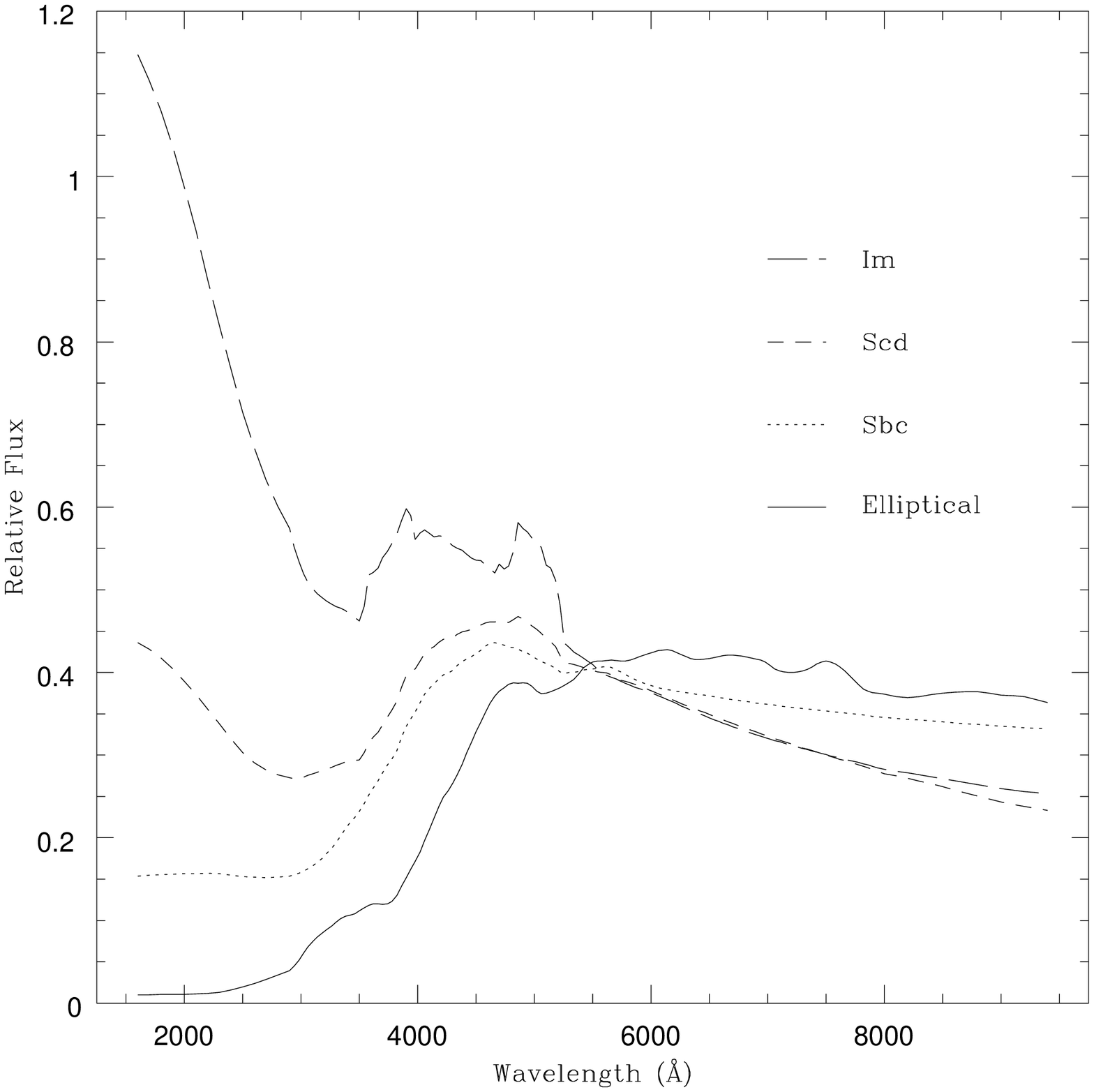]{The template spectra normalized to the 
Elliptical template flux at 5500 \AA\ and smoothed over 100
\AA. \label{spectraNorm}}

\figcaption[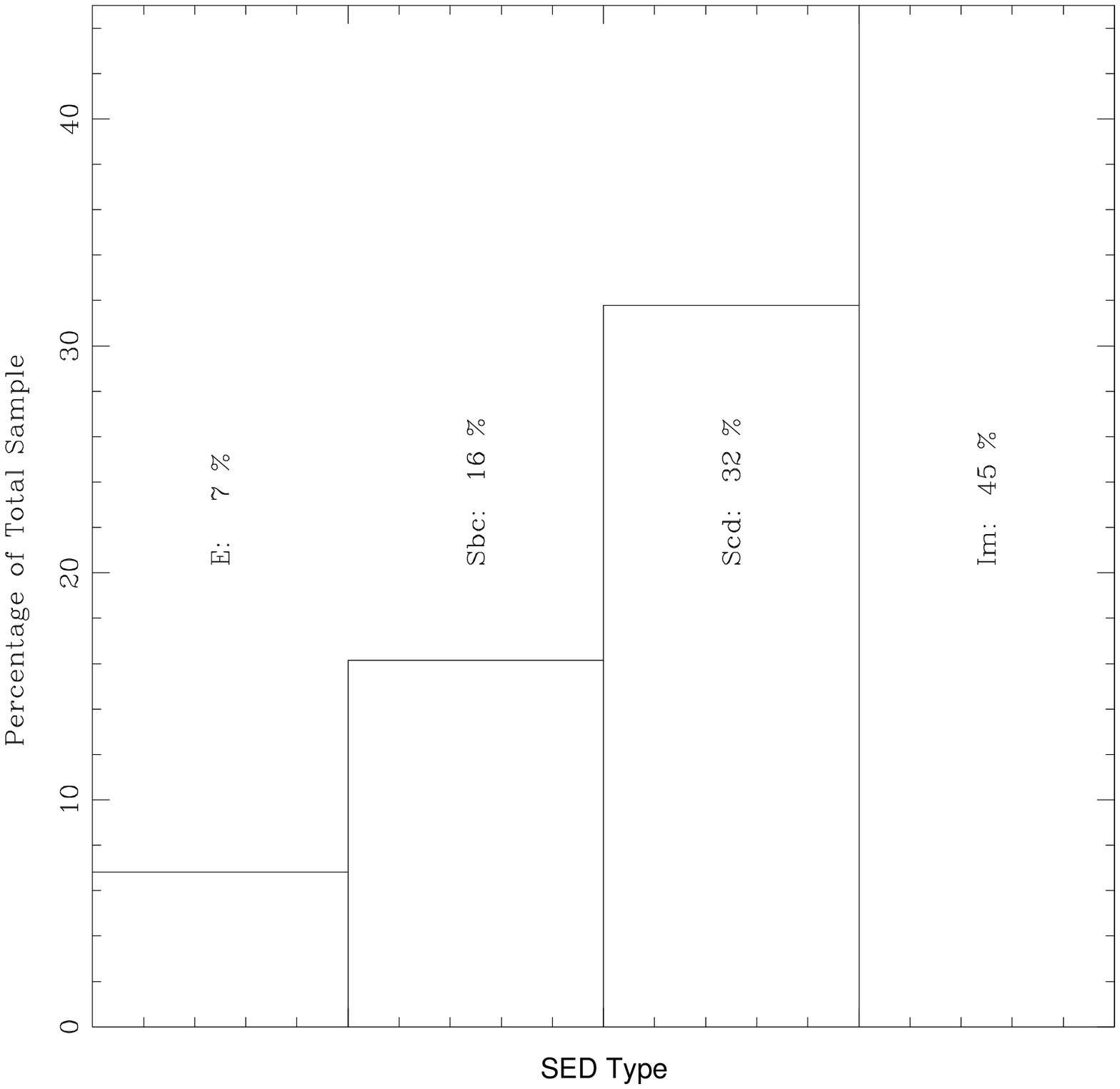]{The SED Percentage of total sample. Note 
that the Elliptical types are diminished relative to the field due to
the four band selection effect.\label{sedPercentage}}

\figcaption[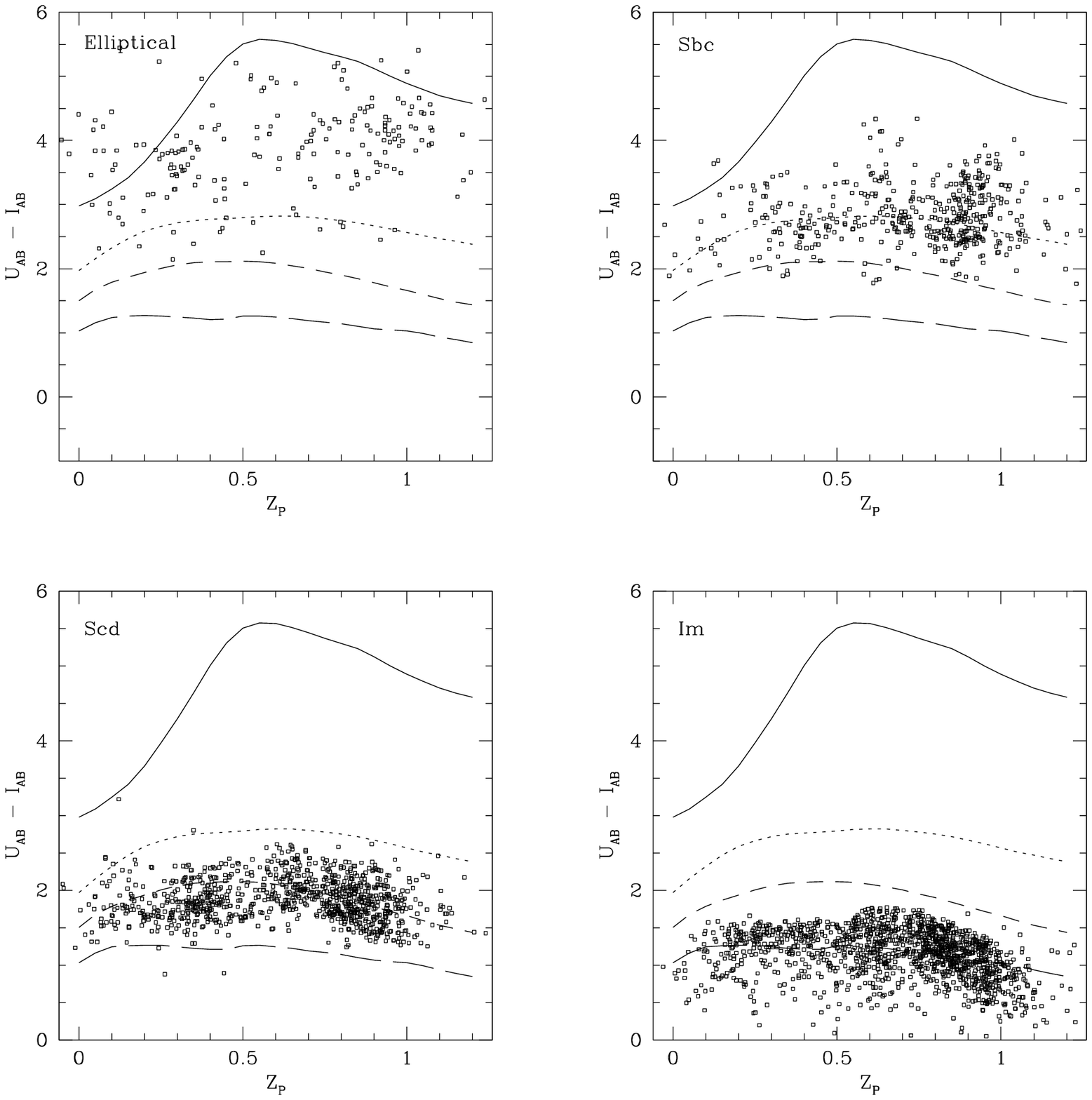]{A comparison between the templates and the 
classified sources in the \U - \I color.\label{tui}}

\figcaption[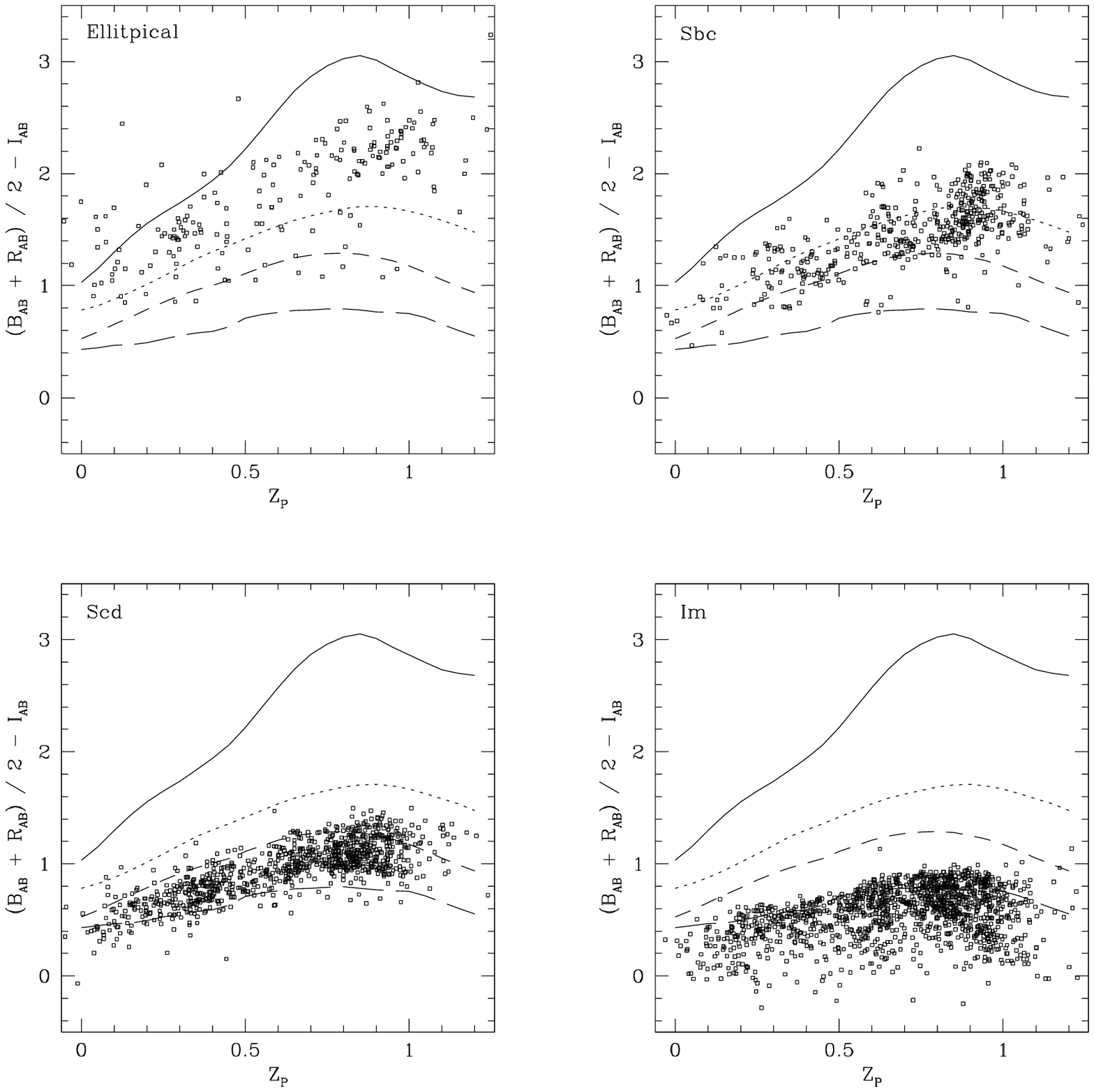]{A comparison between the templates and the 
classified sources in an interpolated \V - \I color.\label{tbi}}

\figcaption[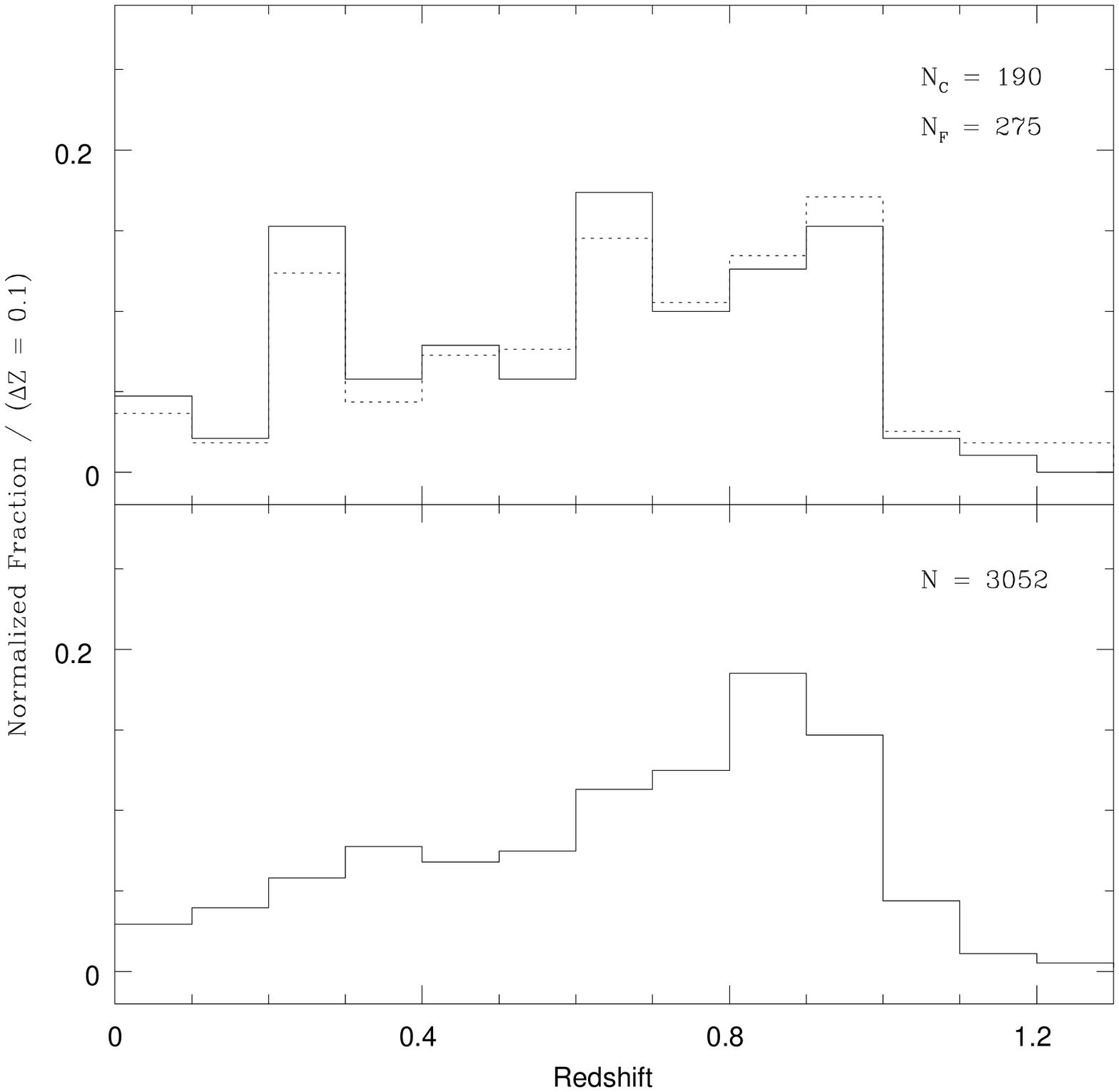]{A comparison between the actual number-redshift 
relation (top) defined by the combined DEEP and CFRS calibrating
redshift sample (solid line) and the full spectroscopic sample (dashed
line), and the number-redshift relation for the full photometric
sample using redshift intervals of 0.1 magnitudes.\label{nOfZ}}

\figcaption[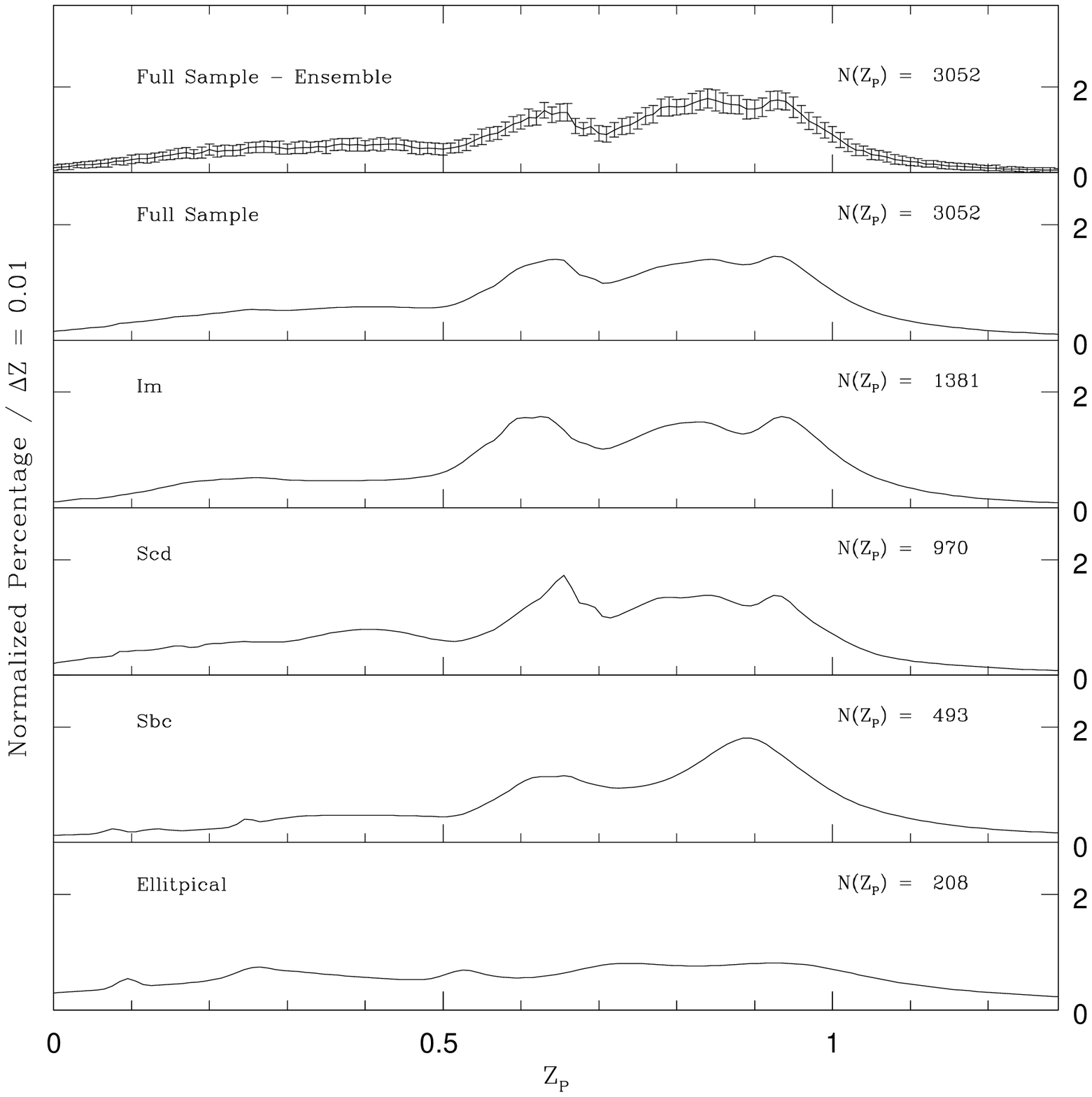]{The number-redshift distribution for both 
the full sample as derived from 100 different ensembles including the
one sigma errors and the full sample derived analytically, as well as
a function of spectral type. Each distribution has been normalized by
the total number of galaxies in the distribution and scaled to
percentages.\label{nOfZTypes}}

\begin{deluxetable}{cccccc}
\tablecaption{The integration times for each of the four filters 
and the corresponding completeness and photometric magnitude limits.
\label{limits}}
\tablewidth{0pt}
\tablehead{
\colhead{Filter}
&\colhead{Integration Time (s)}
&\colhead{$90\%$ }
&\colhead{$50\%$}
&\colhead{$\sigma \approx 0.02$ }
&\colhead{$\sigma \approx 0.1$ }
}

\startdata
U	&40129  &25.94  &26.65	&23.74	&25.92 \nl
B	&16680  &25.26  &26.49	&23.25	&25.64 \nl
R	&4500   &24.45  &25.41	&22.70	&24.66 \nl
I	&4200   &23.97  &24.74	&22.03	&23.94 \nl
\enddata
\end{deluxetable}

\begin{deluxetable}{cccccccccc}
\tablecaption{The completeness limits, turnover point, and measured
slopes for the number-magnitude relationships from our CCD data,
published photographic surveys, and the HDF.
\label{numCountTable}}
\tablewidth{0pt}
\tablehead{
\colhead{Band}
& \colhead{$50 \%$}
& \colhead{$90 \%$}
& \colhead{Turnover}
& \colhead{$\alpha_{Full}$} 
& \colhead{$\alpha_{Low}$}
& \colhead{$\alpha_{High}$}
& \colhead{$\alpha_{Photo}$\tablenotemark{a}}
& \colhead{$\alpha_{HDF}$\tablenotemark{b}}
}

\startdata
$U_{AB}$ &25.94 &26.65 &24.68 	&0.40 &0.51 	&0.24 	&0.6	&0.23 \nl
$B_{AB}$ &25.26 &26.49 &24.45 	&0.39 &0.51 	&0.22 	&0.5	&0.24 \nl
$R_{AB}$ &24.45 &25.41 &---	&0.34 &--- 	&--- 	&0.4	&0.24 \nl
$I_{AB}$ &23.97 &24.74 &---	&0.32 &---	&--- 	&0.4	&0.25 \nl
\nl
\enddata
\tablenotetext{a}{Slopes published by David Koo (1986)} 
\tablenotetext{b}{These slopes are our own fits to published
corrected counts provided by N. Metcalfe (1996). The magnitude range
used was set to minimize the low statistics at the bright end, and the
incompleteness at the faint end.}
\end{deluxetable}

\begin{deluxetable}{ccc}
\tablecaption{The algorithmic details for generating the final photometric
redshift from the initial estimate generated from the global third
order fit. The fit region is the range of global redshift estimates
over which the local fit is applied. The calibration region indicates
the range of calibration redshifts used to constrain the local
fit.\label{algorithmZ}}
\tablehead{
\colhead{Interval Name}
&\colhead{Fit Region} 
&\colhead{Calibration Region} 
}
\startdata
Low		&$z \leq 0.25$		&$(0.0, 0.4]$\nl
Low Break	&$0.25 < z \leq 0.5$	&$(0.2, 0.55]$\nl
Medium		&$0.5 < z \leq 0.7$	&$(0.4, 0.8]$\nl
High Break	&$0.7 < z \leq 0.9$	&$(0.6, 1.0]$\nl
High		&$0.9 < z$		&$(0.8, 1.2]$\nl
\enddata
\end{deluxetable}

\end{document}